\documentclass[twocolumn,times]{aastex631}
\hypersetup{linkcolor=red,citecolor=green,filecolor=cyan,urlcolor=magenta}
\usepackage{amsmath}	% Advanced maths commands

\usepackage{rsfso}%for fancy R

\begin{document}

\title{The mass fallback rate of the debris in relativistic stellar tidal disruption events}

\author[0000-0003-3028-8239]{T. Jankovič}
\affiliation{Center for Astrophysics and Cosmology, University of Nova Gorica, SI-5000 Nova Gorica, Slovenia}
%\correspondingauthor{Taj Jankovič} 
\email{taj.jankovic@ung.si}

\author[0000-0002-0908-914X]{A. Gomboc}
\affiliation{Center for Astrophysics and Cosmology, University of Nova Gorica, SI-5000 Nova Gorica, Slovenia}



	\begin{abstract}
		Highly energetic stellar tidal disruption events (TDEs) provide a way to study black hole characteristics and their environment. We simulate TDEs with the \textsc{Phantom} code in a general relativistic and Newtonian description of a supermassive black hole's gravity. Stars, which are placed on parabolic orbits with different parameters $\beta$, are constructed with the stellar evolution code \textsc{MESA} and therefore have realistic stellar density profiles. We study the mass fallback rate of the debris $\dot{M}$ and its dependence on the $\beta$, stellar mass and age as well as the black hole's spin and the choice of the gravity's description. We calculate peak value $\dot{M}_\mathrm{peak}$, time to the peak $t_\mathrm{peak}$, duration of the super-Eddington phase $t_\mathrm{Edd}$, time $t_{>0.5\dot{M}_\mathrm{peak}}$ during which $\dot{M}>0.5\dot{M}_\mathrm{peak}$, early rise-time $\tau_\mathrm{rise}$ and late-time slope $n_\infty$. We recover the trends of $\dot{M}_\mathrm{peak}$, $t_\mathrm{peak}$, $\tau_\mathrm{rise}$ and $n_\infty$  with $\beta$, stellar mass and age, which were obtained in previous studies. We find that $t_\mathrm{Edd}$, at a fixed $\beta$, scales primarily with the stellar mass, while $t_{>0.5\dot{M}_\mathrm{peak}}$ scales with the compactness of stars. The effect of SMBH's rotation depends on the orientation of its rotational axis relative to the direction of the stellar motion on the initial orbit. Encounters on prograde orbits result in narrower $\dot{M}$ curves with higher $\dot{M}_\mathrm{peak}$, while the opposite occurs for retrograde orbits. We find that disruptions, at the same pericenter distance, are stronger in a relativistic tidal field than in a Newtonian. Therefore, relativistic $\dot{M}$ curves have higher $\dot{M}_\mathrm{peak}$, and shorter $t_\mathrm{peak}$ and $t_\mathrm{Edd}$. 
	\end{abstract}



\section{Introduction} \label{sec:intro}

	The majority of galactic centers contain a supermassive black hole with mass $M_\mathrm{bh}$ in the range of $10^5$--$10^{9.5}\, \mathrm{M_\odot}$, surrounded by a central stellar cluster. There is a probability of about $10^{-5}$--$10^{-4}$ /galaxy/year that a star from such a cluster is scattered and brought in the proximity of the black hole  \citep{Stone2015, Alexander_2017}. The star with mass $M_\star$ and radius $R_\star$ is disrupted if it enters the tidal zone, a region where black hole's tidal forces overcome stellar self-gravity. In the case of a non-rotating SMBH, tidal zone can be approximated as a sphere with a radius equal to the tidal radius $R_\mathrm{t}= R_\star \left( M_\mathrm{bh} /M_\star\right)^{1/3}$. The fate of the stellar debris depends on its total energy. Parts of the debris with positive energy are unbound and escape from the gravitational potential of the black hole. Conversely, debris with negative total energy is bound and returns to the black hole's vicinity, where it can form an accretion disk, which may emit radiation for months to years.

	Currently there are $\approx 100$ observed TDEs \citep{komossa, van_Velzen_2019}. This number is expected to increase drastically ($\approx 1000$ new sources per year) with the start of new wide-field optical surveys (e.g. Vera Rubin observatory) \citep{van_Velzen_2011, Bricman_2020}. In order to distinguish TDEs from other sources with similar locations (galactic center) or light curves, such as active galactic nuclei and supernovae, and to determine TDE parameters, accurate observations and detailed theoretical models are needed.

	In the $80$'s and $90$'s, the first theoretical studies led to the discovery of various mechanisms of the disruption and the debris fallback phase \citep{Carter_1982, rees}. TDEs were also studied with numerical simulations. First simulations were based on affine models \citep{Carter_1985} or on SPH codes \citep{evans}. SPH simulations were severely hindered by a low resolution since the number of particles was several orders of magnitude lower than what is used nowadays. For instance, \citet{evans} used $\sim 10^4$ particles, while current simulations use $\sim 10^6$ particles \citep{liptai2019disc}. Initial studies were mainly focused on the disruption phase, however, as technology advanced with time, studies focused also on later stages --- on the return of the bound debris to the SMBH's vicinity, and the formation and evolution of the accretion disk \citep{Ayal_2000, Hayasaki_2013, bonnerot, liptai2019disc, Clerici_Gomboc_2020}. 
	
	We focus on an analysis of disruptions of stars with realistic stellar profiles in a gravitational potential of a SMBH, which is described in general relativity (relativistic TDEs). A detailed study of relativistic TDEs of realistic main sequence stars was performed by \citet{Ryu_2020a, Ryu_2020b, Ryu_2020c, Ryu_2020d}. Other previous studies have focused on TDEs in a simplified description of SMBH's gravity, such as pseudo or generalized Newtonian potentials, and/or on disruptions of polytropic stars e.g. \citet{Hayasaki_2013, tejeda,Guillochon_2013, Gafton_2019, Golightly_2019, lawsmith2020stellar}. However, both the importance of general relativity and the accurate description of the stellar density profile are expected to play an important role. By simulating relativistic disruptions of realistic stars it would be possible to determine the deviation from previous results, obtained in a Newtonian gravitational potential or with simpler stellar models.
	
	We simulate relativistic disruptions of realistic stars for different stellar and orbital para\-meters, and calculate the mass fallback rate of the debris $\dot{M}$. $\dot{M}$ is often assumed to be directly related to the observed light curve \citep{rees, Guillochon_2013, Law_Smith_2019}. The reason for this assumption is a $\dot{M}$ decay $\propto t^{-5/3}$, a relation analytically derived by \citet{rees} and also observed in TDE light curves \citep{komossa, van_Velzen_2021}. Furthermore, this is supported by several simulations, which indicate short disk circularization times and therefore confirm the tight relation between the fallback rate and the energy dissipated in the disk \citep{bonnerot, liptai2019disc}. 
	
	We compare the results from relativistic disruptions to results from non-relativistic simulations (in a Newtonian description of SMBH's gravity) and characterize the effect of SMBH's rotation on the $\dot{M}$. We also determine the differences between $\dot{M}$ of realistic stars with various ages and masses.

	This paper is organized as follows. In Section \ref{sec2} we describe our method and in Section \ref{sec3} we provide our results. In Section \ref{sec:4} we discuss the results and potential caveats. Section \ref{sec5} summarizes our main conclusions. In Appendix we provide parameters of codes used in our simulations and the resolution test.

	\section{Method} \label{sec2}
	We explore the disruption parameter space by simulating disruptions with different values of the stellar mass $M_\star$, age, parameter $\beta$ and SMBH's spin $a$. Realistic stellar profiles are obtained with the stellar evolu\-tion code \textsc{MESA} \citep{Paxton_2010} and are converted to a 3D particle distribution with the program \textsc{MESA2HYDRO} \citep{Joyce_2019}. The process of disruption is simulated with the version v2021.0.0 of the \textsc{Phantom} software  \citep{price}. 
	
	\subsection{Simulations}
	We present results from 52 simulations of stellar tidal disruption events in a relativistic (GR) and 30 simulations in a non-relativistic (NR), Newtonian, gravitational field of the SMBH. We use a stellar evolution software \textsc{MESA} and construct stars with masses $M_\star=0.6$, 1, 2, 3$\, \mathrm{M_\odot}$ at the beginning of the main sequence (zero age main sequence ---ZAMS) and at the end of the main sequence (terminal age main sequence ---TAMS). ZAMS is defined with the condition that $99\%$ of the total energy is generated in nuclear reactions (fusion of hydrogen into helium), while the condition for TAMS is that the relative abundance of hydrogen in the core is lower than $0.1\%$. Stars with $M_\star = 0.6\, \mathrm{M_\odot}$ have a lifespan longer than the age of the Universe. Therefore, we use a different TAMS condition for these stars --- we stop the stellar evolution $10\, $Gyr after ZAMS. In all cases we assume initial solar metallicity.

	\textsc{MESA} generates a 1D density profile, which has to be converted to a particle distribution in order to be correctly processed by \textsc{Phantom}. For this purpose we use the \textsc{MESA2HYDRO} code \citep{Joyce_2019}. \textsc{MESA2HYDRO} distributes equal mass particles in concentric shells according to the desired density profile. The most interior part is modelled as a sink particle\footnote{A sink particle interacts only gravitationally with the other particles. It is used in the most central stellar region where the density can be several orders of magnitude larger than the density in outer layers. By modelling the stellar center with a single particle, sink particle, it is possible to avoid modelling stellar center with a high number of equal mass particles.}. The full extent of parameters used in $\textsc{MESA}$, $\textsc{MESA2HYDRO}$ and $\textsc{Phantom}$ is presented in Appendix \ref{app:sim}. We generate stellar distributions with $\approx 10^6$ particles and relax them for $\approx 20$ dynamical timescales --- we run these simulations in the absence of an external gravitational field (only stellar self-gravity) in order to dampen velocity perturbations of particles. We also perform a resolution test (see Appendix \ref{app:1}) and find that our choice of the particle number is appropriate.

	Relaxed stars are placed in parabolic orbits at a distance $5R_\mathrm{t}$ from the SMBH with mass $M_\mathrm{bh}=10^6\, \mathrm{M_\odot}$. For GR simulations we calculate the initial positions and velocities according to a relativistic description adapted from the Appendix in \citet{Tejeda_2017}. In the case of NR simulations, we use a combination of Kepler's orbit equation and the energy equation of a parabolic orbit. For each stellar mass, we consider up to 4 different values of the parameter $\beta\equiv R_\mathrm{t}/r_\mathrm{p}=1$, 3, 5, 7, defined as the ratio between the tidal radius $R_\mathrm{t}$ and the pericenter distance $r_\mathrm{p}$. Due to our initial setup, stars with the same age, mass and $\beta$ have the same $r_\mathrm{p}$ in GR and NR. We use an adiabatic equation of state with $\gamma=5/3$, which corresponds to a gas pressure dominated regime. We stop the simulations between the first and second passage --- after the disruption of the star and before the most bound debris returns to the proximity of the SMBH. For the majority of encounters, this corresponds to approximately $55\, $h after the disruption occurs. We note that for higher $\beta$ the second passage happens sooner and simulations were stopped earlier. The effect of SMBH's rotation is taken into account by considering values of the SMBH's spin $a=\pm 0.99$. We use $a>0$ when the star is on a prograde orbit --- when the direction of the initial stellar motion and SMBH's rotation are aligned. Retrograde orbits are defined as $a<0$. For encounters with $\beta=1$ the effect of SMBH's rotation on the process of disruption is negligible. We consider disruptions of stars on non-inclined orbits --- on orbits in the SMBH's equatorial plane. Table \ref{tab:ic} lists the parameter space of simulated disruptions in this work\footnote{We do not consider $\beta=7$ encounters for $M_\star=0.6$, because the orbits are plunging. We simulate TDEs with $a=-0.99$ only for encounters with $\beta=3$ because for the majority of disruptions with a higher $\beta$ a part of the stellar gas is on plunging orbits.}.

		\begin{table*}
		\caption{Parameter space of all the simulations. "+" and "-" symbols correspond to disruptions by a rotating SMBH, with $a=0.99$ and $a=-0.99$, respectively. Ratio between the central density and the average density $\rho_\mathrm{c}/\overline{\rho}$ determines compactness of the star. Values of $r_\mathrm{p}$ correspond to pericenter distances of the star during the first passage.}             % title of Table
		\label{tab:ic}      % is used to refer this table in the text
		\centering                          % used for centering table
		\begin{tabular}{c | c | c | c | c  | c }        % centered columns
			\hline\hline                 % inserts double horizontal lines
			$M_\star\, [\mathrm{M_\odot}]$  & $R_\star \, [\mathrm{R_\odot}]$ & Age$\,[\mathrm{Gyr}]$  & $\rho_\mathrm{c}/\overline{\rho} $ & $\beta$ & $r_\mathrm{p}/R_\mathrm{\odot}$\\    % table heading 
			\hline\hline                      % inserts single horizontal line
			0.6 & 0.56 & 0 & 33 & 1, 3, 5  & 65.6, 21.9, 13.1\\
			& &  &  & $3^{-}$, $3^{+}$, $5^{+}$ &    $18.2^{-}$, $24.0^{+}$, $16.5^{+}$\\
			0.6 & 0.58 & 10 & 43 & 1, 3, 5  & 69.3, 23.1, 13.9\\
			& &  &  & $3^{-}$, $3^{+}$, $5^{+}$ &    $19.8^{-}$, $25.2^{+}$, $17.1^{+}$\\
			1 & 0.86 & 0 & 56 & 1, 3, 5, 7 & 86.9, 29.3, 17.7, 12.8\\
			& &  &  & $3^{-}$, $3^{+}$, $5^{+}$, $7^{+}$ &    $27.0^{-}$, $31.1^{+}$, $20.3^{+}$, $16.3^{+}$\\
			1 & 1.23 & 8.3 & 1363 & 1, 3, 5, 7  & 123.2, 40.7, 24.3, 17.2\\
			& &  &  & $3^{-}$, $3^{+}$, $5^{+}$, $7^{+}$ &    $39.1^{-}$, $42.1^{+}$, $26.3^{+}$, $19.9^{+}$\\ 
			2 & 1.27 & 0 & 96 & 1, 3, 5, 7  & 100.0, 32.8, 19.4, 13.6\\
			& &  &  & $3^{-}$, $3^{+}$, $5^{+}$, $7^{+}$ &    $30.7^{-}$, $34.9^{+}$, $21.8^{+}$, $16.8^{+}$\\
			2 & 2.54 & 1.0 & 2426 & 1, 3, 5, 7  & 201.7, 68.3, 41.5, 29.9\\
			& &  &  & $3^{-}$, $3^{+}$, $5^{+}$, $7^{+}$ &    $67.3^{-}$, $69.3^{+}$, $42.8^{+}$, $31.6^{+}$\\
			3 & 1.57 & 0 & 58 & 1, 3, 5, 7  & 109.1, 36.2, 21.7, 15.5\\
			& &  &  & $3^{-}$, $3^{+}$, $5^{+}$, $7^{+}$ &    $34.4^{-}$, $37.7^{+}$, $23.9^{+}$, $18.3^{+}$\\
			3 & 3.25 & 0.3 & 1908 & 1, 3, 5, 7 & 222.0, 75.5, 45.8, 32.9\\
			& &  &  & $3^{-}$, $3^{+}$, $5^{+}$, $7^{+}$ &    $74.5^{-}$, $76.4^{+}$, $47.0^{+}$, $34.5^{+}$\\
			\hline                                   %inserts single line
		\end{tabular}
	\end{table*}

	\subsection{Postprocessing}
	
	When a star enters SMBH's tidal field it is squeezed in the direction orthogonal to the orbital plane and in the radial direction due to the tidal field. This increases the stellar internal pressure. When the acceleration due to the gas pressure overcomes the acceleration due to the tidal squeezing, the collapsing layers bounce back, and the disruption occurs. The SMBH's tidal field affects the debris's energy distribution, which we use for calculations of the mass fallback rate of the debris --- we adopt a similar procedure as \citet{Lodato_2009,Guillochon_2013,Gafton_2019,lawsmith2020stellar, Ryu_2020b}. 
	
	We calculate the total specific energy (expressed in units of the particle mass) of the debris in non-relativistic encounters as  
	\begin{equation} \label{eq:Enr}
		\epsilon=\epsilon_\mathrm{kin}+\epsilon_\mathrm{g}=\frac{1}{2}v^2 - \frac{GM_\mathrm{bh}}{r}.
	\end{equation}
	Here $G$ is the gravitational constant, $r$ is the radial distance to the SMBH and $v$ is the size of the velocity vector. 
	
	In relativistic encounters we calculate $\epsilon$ in Boyer-Lindquist (BL) coordinates as in the Appendix in \citet{Tejeda_2017}. In units $G=c=1$ the expression for the total specific energy is
	\begin{equation}\label{eq:Egr}
		\epsilon=\Gamma \left(  1 - \frac{2M_\mathrm{bh}r}{\rho^2}  \left[  1 - a\left(  \frac{x\dot{y} - y\dot{x}}{r^2 +a^2} \right)   \right] \right),
	\end{equation}
	where $\rho^2=r^2 +\frac{a^2 z^2}{r^2}$ . The position is defined with $x$, $y$, $z$ coordinates, which are calculated in the reference frame of the SMBH. $\Gamma$ is the Lorentz factor expressed in BL coordinates as

	\begin{equation}
		\begin{split}
			\Gamma =\frac{\mathrm{d}t}{\mathrm{d}\tau}=
			&\bigg(1-\dot{x}^2-\dot{y}^2-\dot{z}^2- \\
			&  -\frac{2M_\mathrm{bh}r}{\rho^2} \left( \left[  1 - a\left( \frac{x\dot{y}-y\dot{x}}{r^2+a^2}\right) \right]^2 +  \right. \\
			&\quad  \left. {}+ \frac{[r^2 (x\dot{x}+y\dot{y}) +(r^2+a^2)z\dot{z}]^2}{r^2 \Delta (r^2 +a^2)} \right) \vphantom{1}  \bigg)^{-1/2}
		\end{split}
	\end{equation}
	Here $\Delta$ is $\Delta=r^2 -2M_\mathrm{bh}r +a^2$.
	
	We use the Equations (\ref{eq:Enr}) and (\ref{eq:Egr}) (expressed in SI units) to calculate $\epsilon$, and the third Kepler's law to determine $\dot{M}$ as\footnote{We have also tried using the same approach as in \citet{Gafton_2019} and calculated $\dot{M}$ from the geodesics. We found negligible differences between the two methods.}
	\begin{equation} \label{eq4}
		\dot{M}=\frac{\mathrm{d}M}{\mathrm{d}t}=\frac{\mathrm{d}M\mathrm{d}\epsilon}{\mathrm{d}\epsilon\mathrm{d}t}=\frac{\mathrm{d}M}{\mathrm{d}\epsilon} \frac{1}{3}\left(2\pi GM_\mathrm{bh} \right)^ {2 / 3}t^{-5/3}.
	\end{equation}
	We obtain the mass distribution of the debris over total energy $ {\mathrm{d}M}/{\mathrm{d}\epsilon}$ from the final snapshot of simulations. We use the iterative approach described in \citet{Guillochon_2013} to calculate the self-bound particles\footnote{In less disruptive encounters, e.g. with a low parameter $\beta$ and/or disruptions of stars with a high central density, it is possible that the stellar core survives. We refer to particles that are bound to the stellar core (and not to the SMBH) as self-bound particles.} and exclude them from calculations of $ {\mathrm{d}M}/{\mathrm{d}\epsilon}$. We fit $ {\mathrm{d}M}/{\mathrm{d}\epsilon}$ distributions with B-spline functions in order to produce smooth $\dot{M}$ curves. We note, that using the Equation (\ref{eq4}) to calculate $\dot{M}$ assumes that the properties of gas are "frozen-in", and the gas is moving on ballistic trajectories. This is not necessarily true if the stellar core survives the disruption and its gravitational potential breaks the "frozen-in" approximation. To check the validity of this approximation, we calculated $\dot{M}$ from snapshots at different times after the disruption (all before the most bound debris returns to the SMBH's vicinity). We find, that the differences between them are negligible
 and conclude, that this approximation is valid also for disruptions with a surviving stellar core.
	
	This approach enables the calculation of characteristic values of the $\dot{M}$ curves:
	\begin{itemize}
		\item the value of the peak $\dot{M}_\mathrm{peak}$,
		\item time $t_\mathrm{peak}$ (from disruption until $\dot{M}=\dot{M}_\mathrm{peak}$),
		\item duration of the super-Eddington phase $t_\mathrm{Edd}$ (during which $\dot{M}>\dot{M}_\mathrm{Edd}$),
		\item duration $t_{>0.5\dot{M}_\mathrm{peak}}$ (during which $\dot{M}>0.5\dot{M}_\mathrm{peak}$),
		\item characteristic rise-time $\tau_\mathrm{rise}$,
		\item late-time slope $n_\infty$.
	\end{itemize}
	 $\tau_\mathrm{rise}$ is calculated by fitting a Gaussian function $\propto e^{-(t-t_\mathrm{peak})^2/(2\tau_\mathrm{rise}^2)}$ to the mass fallback rate of the debris $\dot{M}(t)$ (motivated by  \citet{van_Velzen_2019}) in a range between $0.2\dot{M}_\mathrm{peak}\leq\dot{M}(t)\leq0.8\dot{M}_\mathrm{peak}$ for $t<t_\mathrm{peak}$. $n_\infty$ is obtained by fitting a power-law function $\propto t^{n_\infty}$ to the late-time portion of  $\dot{M}(t)$ curves with no apparent breaking in the power-law dependency. Typically, this corresponds to a range between $\dot{M}(t) \lessapprox 0.1\dot{M}_\mathrm{peak}$ for $t>t_\mathrm{peak}$.
	
	For NR encounters $\dot{M}_\mathrm{peak}$, $t_\mathrm{peak}$ and $t_\mathrm{Edd}$ can be scaled with the stellar mass and radius as
	\begin{flalign}
		\dot{M}_\mathrm{peak} &\propto M_\star^2 R_\star^{-3/2},&& \label{eq_dotMpeak}\\
		t_\mathrm{peak} &\propto M_\star^{-1}R_\star^{3/2}, \label{eq_tpeak}&& \\
		t_\mathrm{Edd} & \propto M_\star^{1/5}R_\star^{3/5}, \label{eq_tedd}&&
	\end{flalign}
	where a constant SMBH mass and a flat $\mathrm{d}M/\mathrm{d}\epsilon$ distribution are assumed \citep{Stone_2019, lawsmith2020stellar}. We also note that a star on the main sequence roughly follows a mass-radius relation $R_\star \propto M_\star^{1}$ and $R_\star \propto M_\star^{0.5}$, for $M_\star \lessapprox 1.5\, \mathrm{M_\odot}$ and $M_\star \gtrapprox 1.5\, \mathrm{M_\odot}$, respectively \citep{book_sa2}. The scaling exponent is higher for lower mass stars due to the increasing role of the convective envelope. The mass-radius relation is also affected by the stellar age --- the scaling exponent is higher for TAMS than for ZAMS stars \citep{m_r_paper_1991}.
	
	\section{Results} \label{sec3}
	
	The mass fallback rate curves $\dot{M}$ are determined by the mass distribution of the debris after the first passage. The mass distribution is affected by the strength of the encounter, stellar structure and SMBH's rotation. Therefore, each of these leave an imprint on the $\dot{M}$ curves and it's characteristic properties.
	Throughout the rest of the paper we divide our analysis in 3 different categories: 
	\begin{itemize}
		\item comparison between GR and NR simulations,
		\item effects of the initial orbit and stellar properties,
		\item effects of the SMBH's rotation.
	\end{itemize}
	For each of these categories we study the effect on the characteristic properties of $\dot{M}$.

	\subsection{The disruption phase}
	
	\subsubsection{First passage}
\begin{figure} [htb!]
		\centering
		\includegraphics[width=0.48\textwidth]{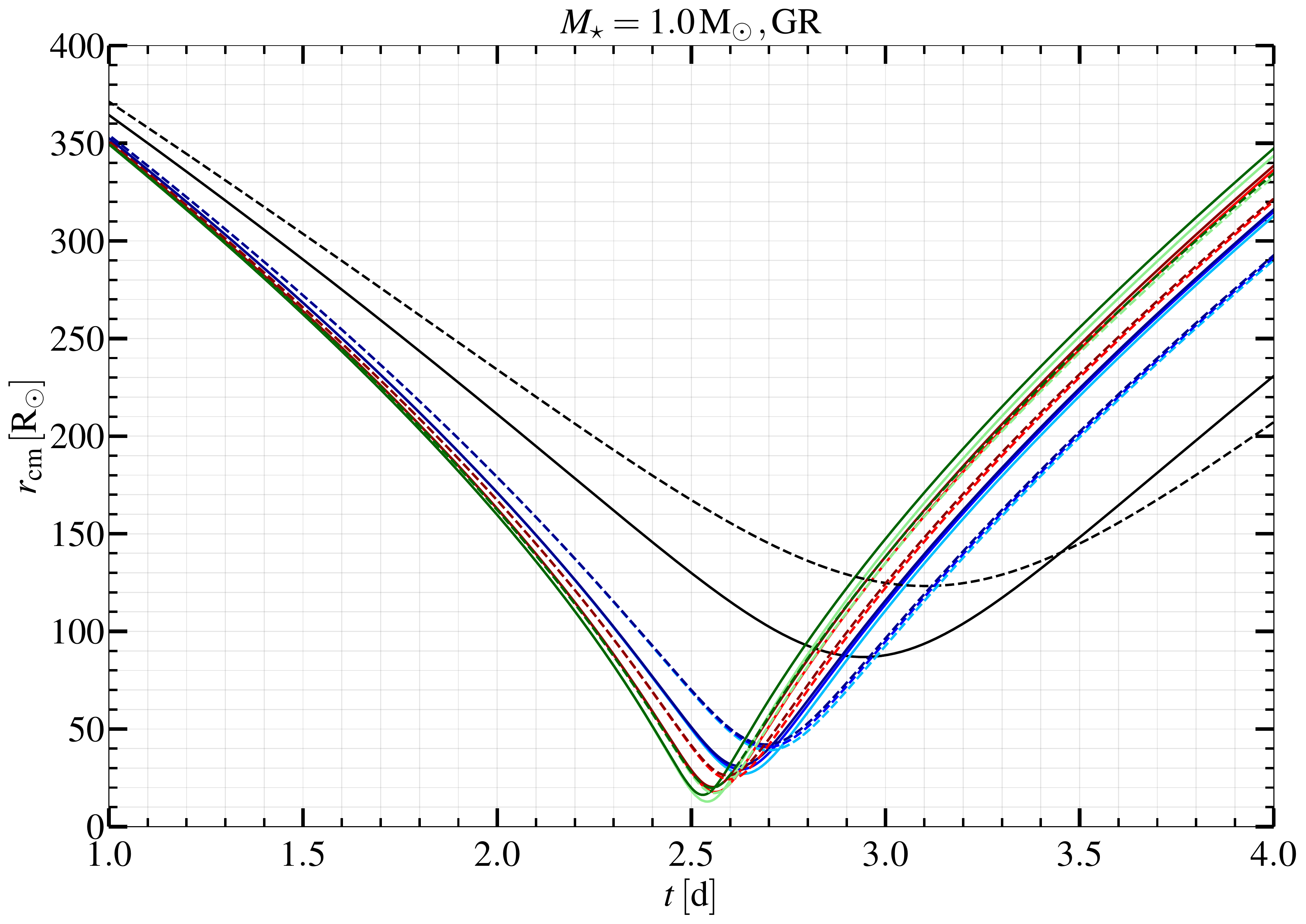}
		\vspace*{0.1cm}
		\begin{minipage}[b]{\linewidth}
			\centering
			\includegraphics[width=\textwidth]{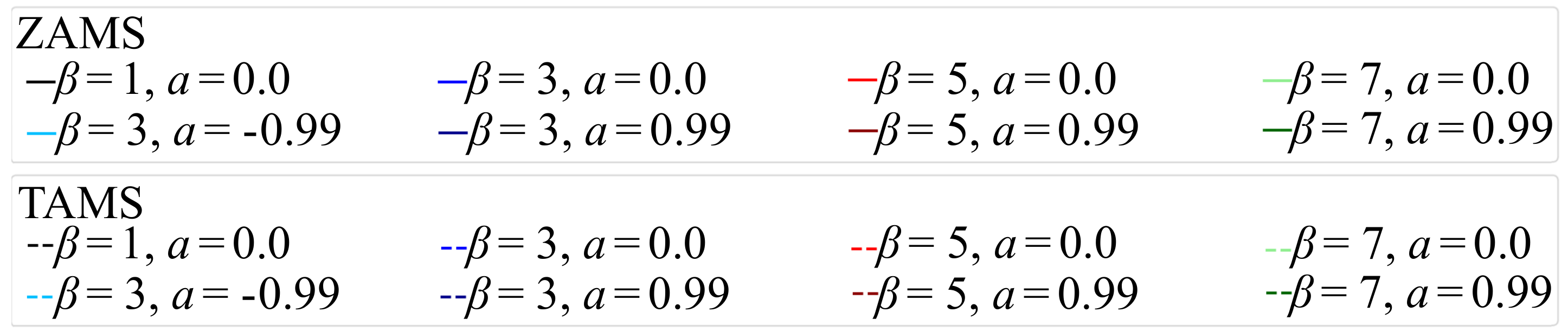}
		\end{minipage}
		\vspace*{-0.1cm}
		
		\caption{Common center of mass distances $r_\mathrm{cm}$ during the first passage for $1\, \mathrm{M_\odot}$ stars for different ages, $\beta$ and $a$. Solid and dashed lines are used for $r_\mathrm{cm}$ of ZAMS and TAMS stars, respectively.}
		\label{s:2}
	\end{figure}

	Figure \ref{s:2} shows common center of mass distances $r_\mathrm{cm}$ during the first passage of $1\, \mathrm{M_\odot}$ stars for different ages, parameters $\beta$ and SMBH's spins $a$. Minima of $r_\mathrm{cm}$ correspond to pericenter distances. Older stars (TAMS) have larger radii and therefore also larger tidal radii, which results in  higher values of $r_\mathrm{p}$. The effect of spin is visible albeit not very strong and the effect differs if stars are on prograde or retrograde orbits due to the relativistic spin-dependent pericenter precession. For the same value of $\beta$ stars on retrograde orbits get closer to the SMBH, while the opposite occurs for stars on prograde orbits. 
	
\begin{figure*} [htb!]
		\centering
		\begin{minipage}[b]{.49\linewidth}
			\includegraphics[width=\textwidth]{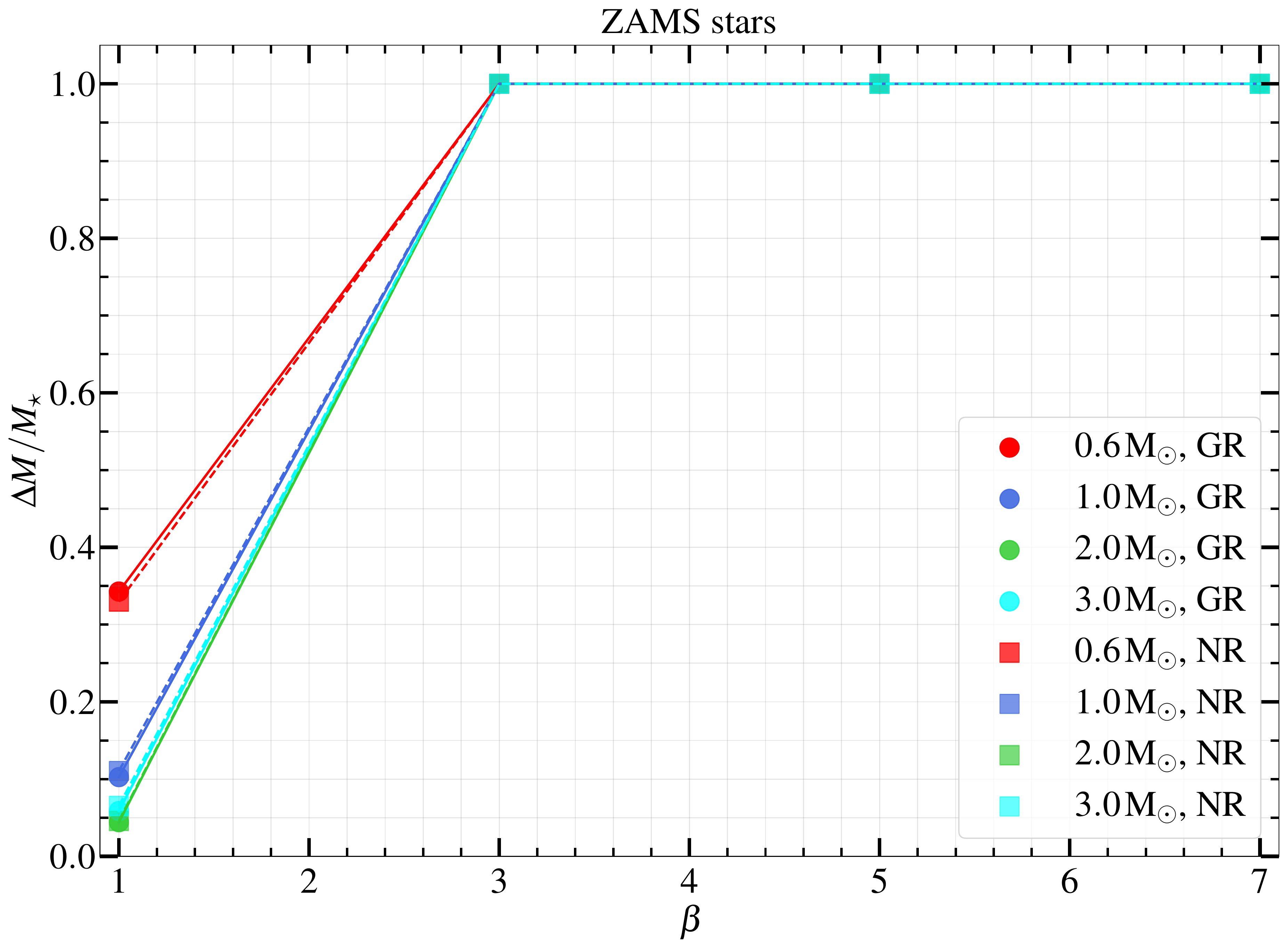}
		\end{minipage}\hfill
		\begin{minipage}[b]{.49\linewidth}
			\includegraphics[width=\textwidth]{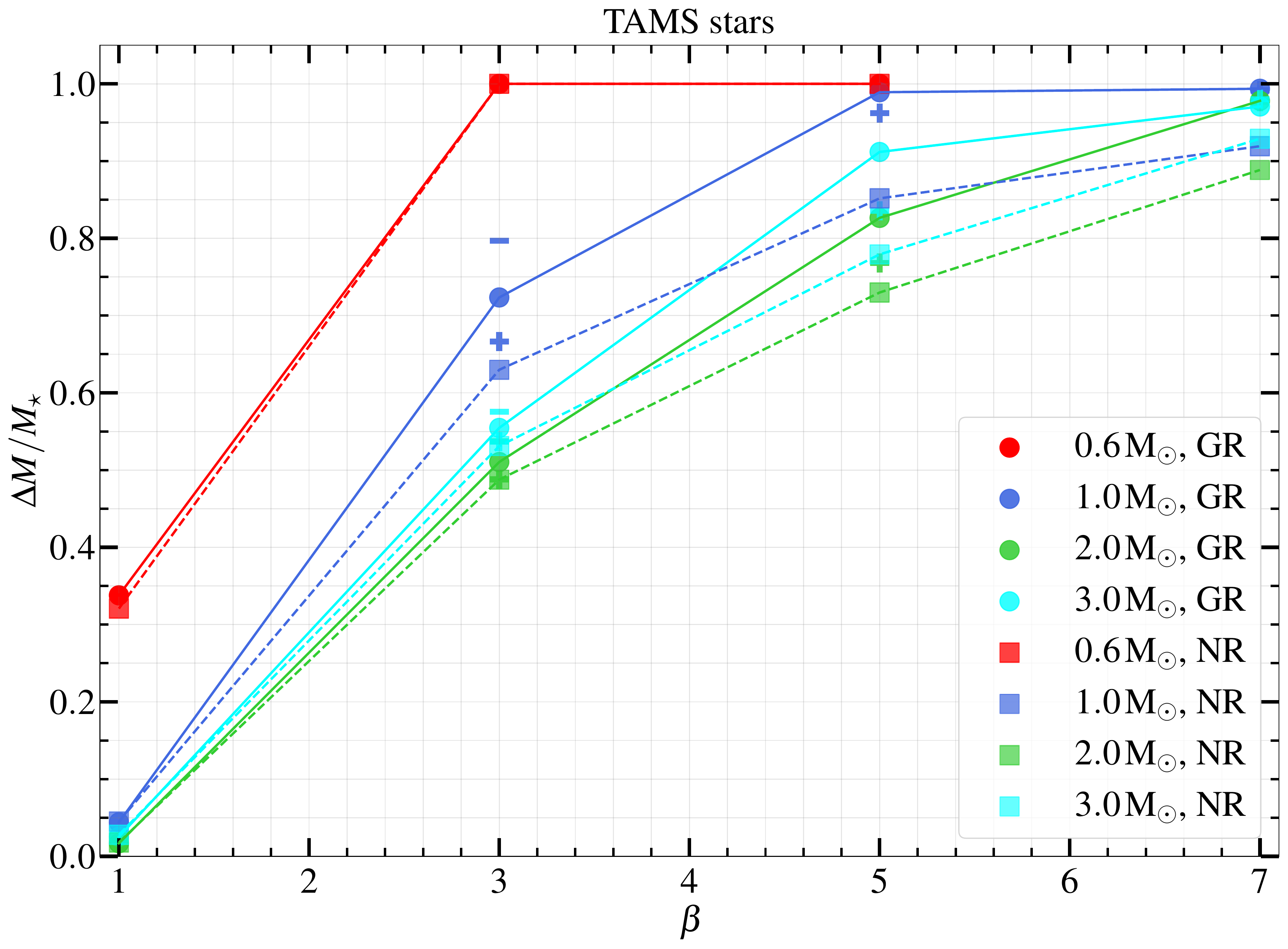}
		\end{minipage}
		\caption{Fraction of the initial stellar mass that is not bound to the star for ZAMS (left) and TAMS (right) stars for different $M_\star$, $a$ and $\beta$. "+" and "-" symbols indicate results from disruptions of stars on prograde and retrograde orbits, respectively. Results for GR and NR simulations are indicated with "$\bullet$" and "$\blacksquare$" symbols, respectively. Values of $\Delta M/M_\star \approx 1$ correspond to total disruptions.}
		\label{s:3}
	\end{figure*}

	TDEs can be divided into 2 groups based on the amount of mass that remains bound to the stellar core after the disruption: partial (PTDEs) and total disruptions (TTDEs). These groups have different $\dot{M}$ curves \citep{Guillochon_2013, Gafton_2019}. For that purpose we define the \emph{lost mass} $\Delta M/M_\star$, where $\Delta M = M_\star - M_\mathrm{self-bound}$. Lost mass is a fraction of the total stellar mass that is not self-bound (bound to the possibly surviving stellar core) --- a fraction that is "lost" to the SMBH \citep{Guillochon_2013}. Figure \ref{s:3} illustrates dependence of the mass lost for all simulations. $\Delta M/M_\star$ increases with $\beta$ and approaches a value of 1 when a TTDE occurs. We see that disruptions of ZAMS stars result in TTDEs (except for $\beta=1$), while disruptions of older TAMS stars result in PTDEs (except for $\beta = 7$). This is partially a consequence of the definition of $\beta\equiv R_\mathrm{t}/r_\mathrm{p}$, where tidal radius depends on the size of the star. Furthermore, older stars have significantly more compact cores (see the ratio $\rho_\mathrm{c}/\overline{\rho}$ in Table \ref{tab:ic}\footnote{In general the value of $\rho_\mathrm{c}/\overline{\rho} $ increases with the stellar mass and age. However, this trend reverses during the transition from a $2\, \mathrm{M_\odot}$ to a $3\, \mathrm{M_\odot}$ star. This is in agreement with \citet{lawsmith2020stellar} who observe a similar transition from $1.5\, \mathrm{M_\odot}$ to a $3\, \mathrm{M_\odot}$ star. The transition is a consequence of a formation of convective core for stars with $M_\star>1.2\, \mathrm{M_\odot}$ \citep{book_sa2}. In a stellar core convective mixing flattens the density profile.}) and therefore require a stronger encounter for a total disruption. On the other hand, stars with lower masses have a smaller tidal radius and are also more easily disrupted due to a less compact core. An exception are $3\, \mathrm{M_\odot}$ stars, that have lower values of $\rho_\mathrm{c}/\overline{\rho}$ than $2\, \mathrm{M_\odot}$ stars (see Table \ref{tab:ic}). Therefore, $2\, \mathrm{M_\odot}$ stars are totally disrupted at a larger $\beta$ than $3\, \mathrm{M_\odot}$ stars.
	
	In general, relativistic disruptions result in a larger mass lost $\Delta M/M_\star$ for the same $\beta$ compared to non-relativistic. We contribute this to a stronger tidal field in GR than in NR, and discuss this effect in Section \ref{sec:4}. The amount of lost mass is also affected by SMBH's rotation. A star on a retrograde orbit has a shorter pericenter distance and experiences a stronger tidal field, which results in a higher lost mass. On the other hand, the amount of lost mass is lower for disruptions of stars on prograde orbits.
	
	\subsubsection{Energy spread}
	
\begin{figure} [htb!]
		\centering
		\includegraphics[width=0.475\textwidth]{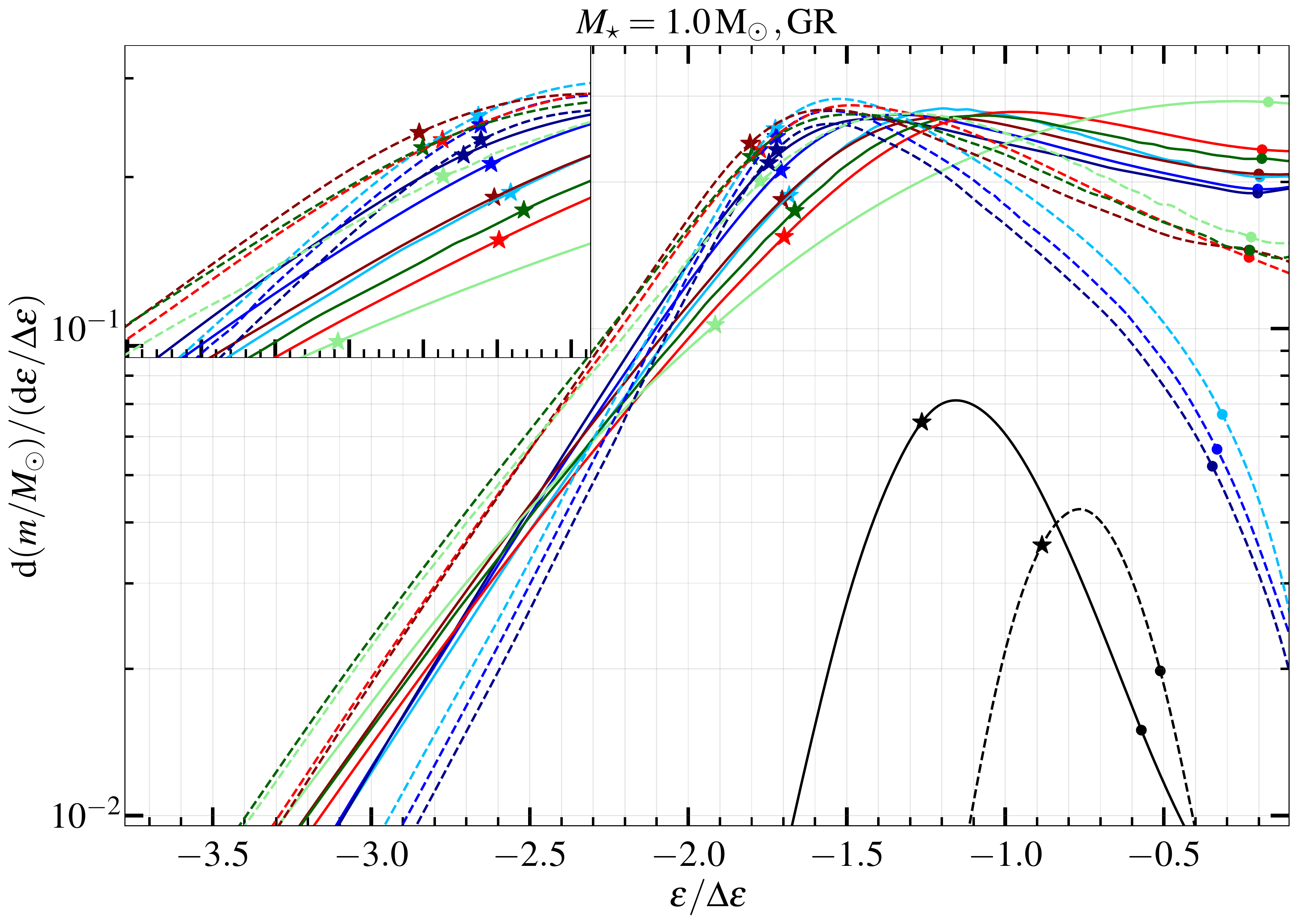}
			\vspace*{0.1cm}
		\begin{minipage}[b]{\linewidth}
			\centering
			\includegraphics[width=\textwidth]{dotM_legend_horizontal_crop.pdf}
		\end{minipage}
		\vspace*{-0.1cm}
		
		\caption{Mass distribution of the debris  over total specific energy $\mathrm{d}M/\mathrm{d}\epsilon$ for $1\, \mathrm{M_\odot}$ calculated from the final snapshot of simulations. "$\star$" symbols illustrate the values of $\mathrm{d}M/\mathrm{d}\epsilon$ that correspond to the maxima of $\dot{M}$, while "$\bullet$" symbols represent the time when $\dot{M}$ decreases below $\dot{M}_\mathrm{Edd}$ (see Figure \ref{fig:dotM}). Solid and dashed lines correspond to results from disruptions of ZAMS and TAMS stars, respectively. Values of $\epsilon$ are normalized by $\Delta \epsilon = GM_\mathrm{bh}R_\star /R_\mathrm{t}^2$.}
		\label{s:4}
	\end{figure}

\begin{figure}
		\centering
		\includegraphics[width=0.475\textwidth]{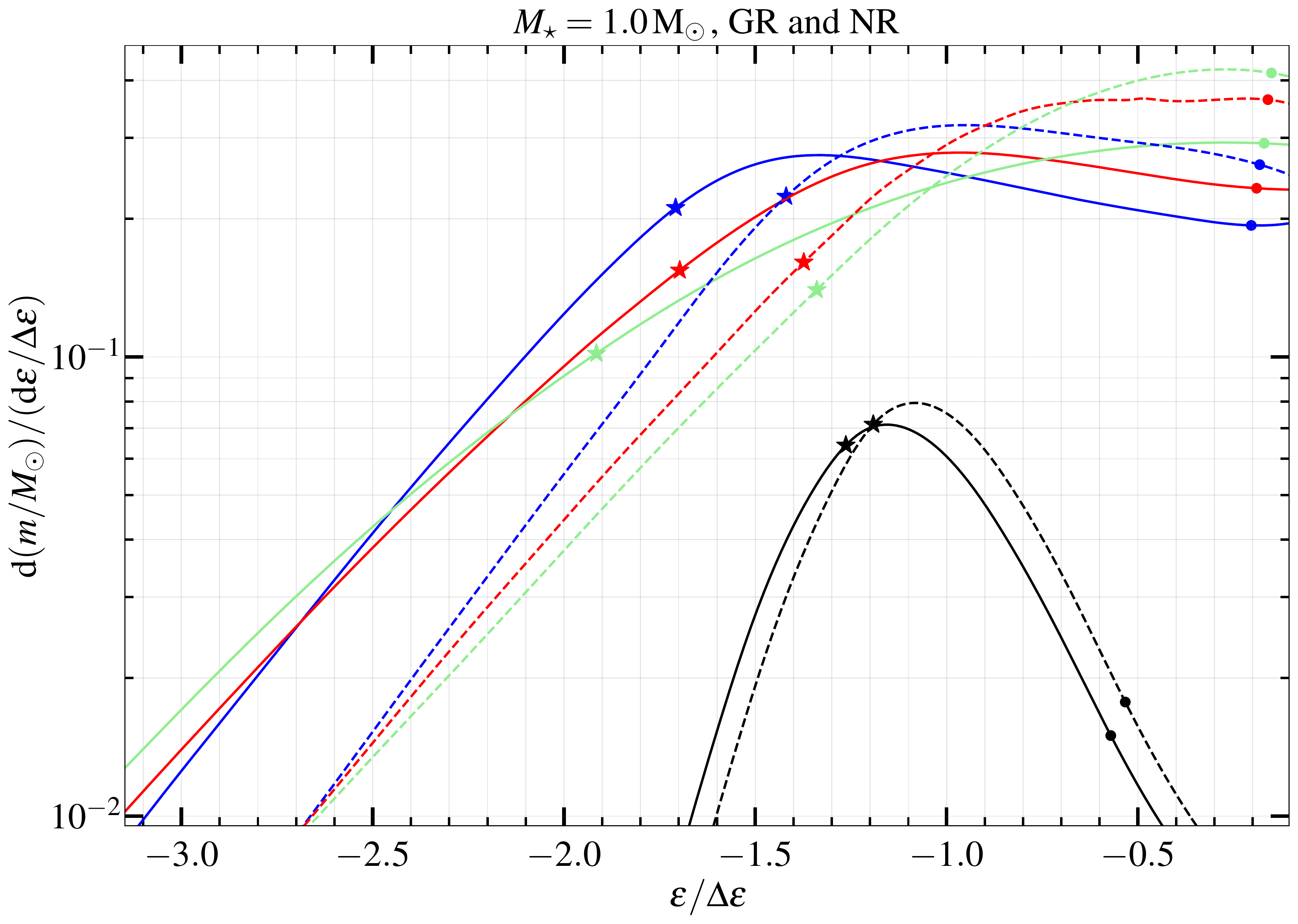}
			\vspace*{0.1cm}
		\begin{minipage}[b]{\linewidth}
			\centering
			\includegraphics[width=\textwidth]{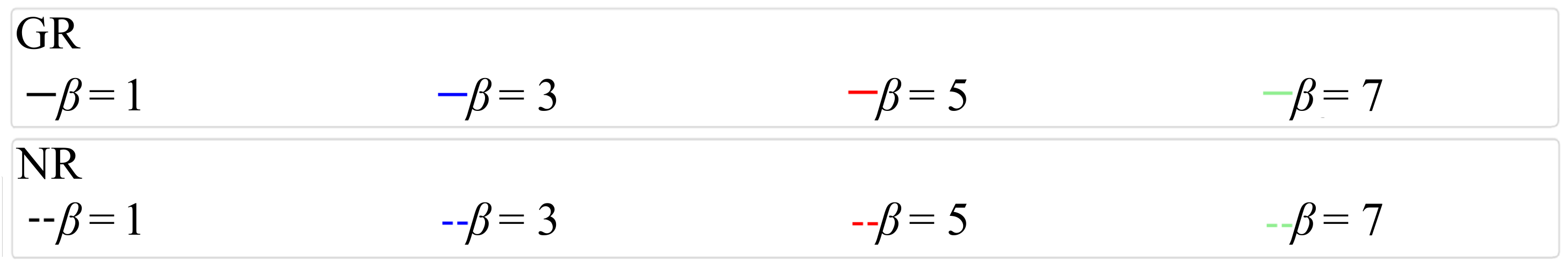}
		\end{minipage}
		\vspace*{-0.1cm}
		
		\caption{Mass distribution of the debris over total specific energy $\mathrm{d}M/\mathrm{d}\epsilon$ for $1\, \mathrm{M_\odot}$ ZAMS stars calculated from the final snapshot of simulations. "$\star$" symbols illustrate the values of $\mathrm{d}M/\mathrm{d}\epsilon$ that correspond to the maxima of $\dot{M}$, while "$\bullet$" symbols represent the time when $\dot{M}$ decreases below $\dot{M}_\mathrm{Edd}$ (see Figure \ref{fig:dotM}). Solid and dashed lines correspond to results from GR and NR disruptions, respectively. Values of $\epsilon$ are normalized by $\Delta \epsilon = GM_\mathrm{bh}R_\star /R_\mathrm{t}^2$.}
		\label{fig:gr_vs_nr}
	\end{figure}

	The effect of tidal forces during the first passage results in an energy spread of the debris which in turn affects the $\dot{M}$ curves. Mass distributions over total specific energy $\mathrm{d}M/\mathrm{d}\epsilon$ for $1\, \mathrm{M_\odot}$ simulations are shown in Figures \ref{s:4} and \ref{fig:gr_vs_nr}. We show the distributions only for $\epsilon<0$, where we do not include gas in the possibly surviving stellar core, because we are interested in the debris bound to the SMBH. We note, that distributions are symmetric with respect to the $y$-axis at $\epsilon=0$ if the unbound gas is included. We mark the values of $\mathrm{d}M/\mathrm{d}\epsilon$ which correspond to the maxima of $\dot{M}$ and to the time, when $\dot{M}$ decrease below the Eddington limit (see Figure \ref{fig:dotM}). We note that maxima of $\dot{M}$ do not correspond to the maxima in $\mathrm{d}M/\mathrm{d}\epsilon$ distributions, since $\dot{M}$ has also the term $\mathrm{d}\epsilon/\mathrm{d}t$ (see Equation \ref{eq4}). Although monotonically decreasing, the $\mathrm{d}\epsilon/\mathrm{d}t$ term shifts the peak values of $\dot{M}$ curves towards more negative values of $\epsilon$. 
	
	For PTDEs the distributions have a pronounced peak of $\mathrm{d}M/\mathrm{d}\epsilon$ as seen in Figure \ref{s:4}. Due to the exclusion of self-bound particles, values of $\mathrm{d}M/\mathrm{d}\epsilon$  sharply decrease as $\epsilon$ approaches 0. With increasing $\beta$ the amount of matter bound to the stellar core is decreasing and the area integral of distributions increases until a TTDE occurs. Disruptions with higher $\beta$ are stronger and the energy spread increases, which causes the debris to occupy a wider range of elliptical orbits after the disruption. The maxima of $\dot{M}$ are shifted towards lower values of $\epsilon$ as the strength of encounters increases and a larger fraction of the debris is moving on more bound elliptic orbits\footnote{We note that in several high $\beta$ disruptions the iterative approach, used to determine self-bound particles, yielded a self-bound mass within $\approx 0.05M_\star$ even if there was no formation of a core-like structure. In those cases we assume a total disruption and extrapolate the last $\approx 10\%$ of $ {\mathrm{d}M}/{\mathrm{d}\epsilon}$.}. For encounters with higher $\beta$ the fallback rate of the debris becomes sub-Eddington at higher energies $\epsilon$ and therefore later times. Disruptions in GR are stronger than in NR, which results in wider energy distributions than NR (see Figure \ref{fig:gr_vs_nr}). At a fixed $\beta$ the steepness of distributions at the low energy is similar in NR and GR, contrary to the energy distribution steepness at low energy for different $\beta$, where the steepness decreases with $\beta$.

	SMBH's spin affects distributions of $\mathrm{d}M/\mathrm{d}\epsilon$ for prograde encounters in a similar way as decreasing $\beta$. For instance, $\mathrm{d}M/\mathrm{d}\epsilon$ distribution for a disruption of a ZAMS star with $\beta=7$, $a=0.99$ is more similar to $\beta=5$, $a=0$ than $\beta=7$, $a=0.$ The opposite occurs for stars on retrograde orbits.

\begin{figure*} [htb!]
		\centering
				\hspace*{0.0005\textwidth}
		\begin{minipage}[b]{0.903\linewidth}
		\includegraphics[width=\textwidth]{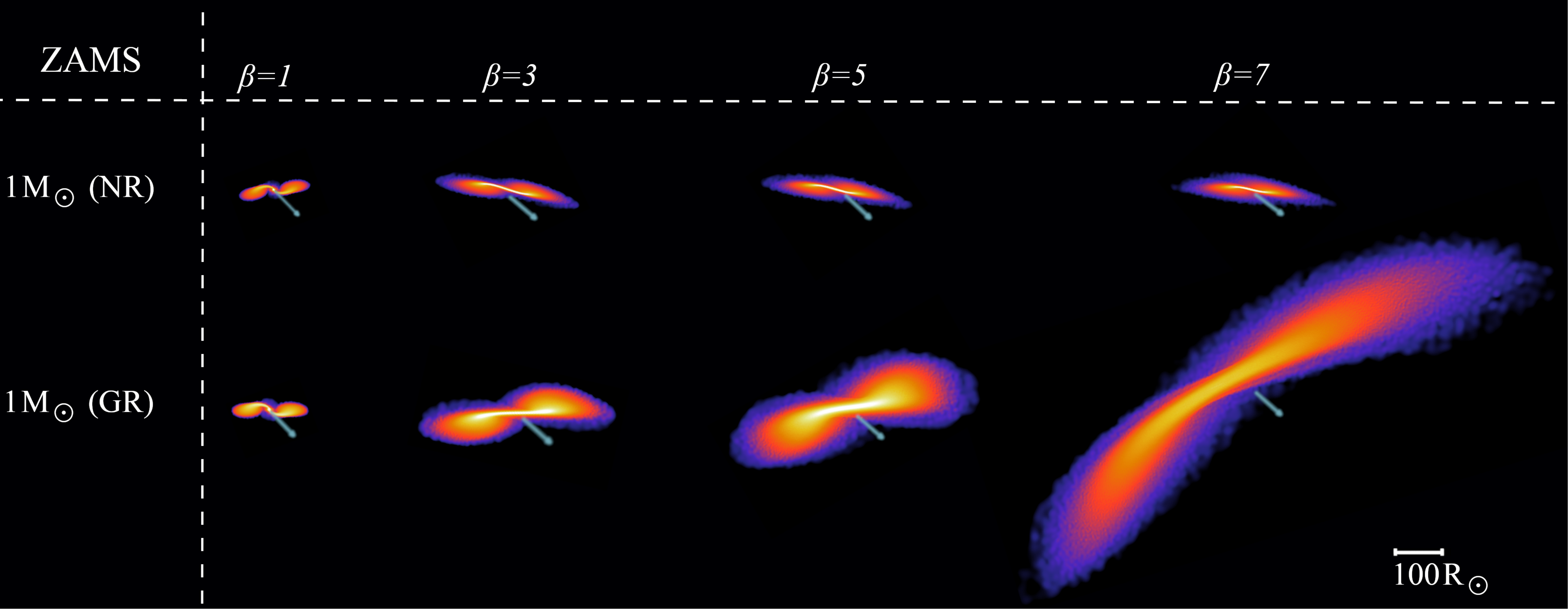}
		\end{minipage}
		\vspace{0.01cm}
			
		\begin{minipage}[b]{0.903\linewidth}
		\includegraphics[width=\textwidth]{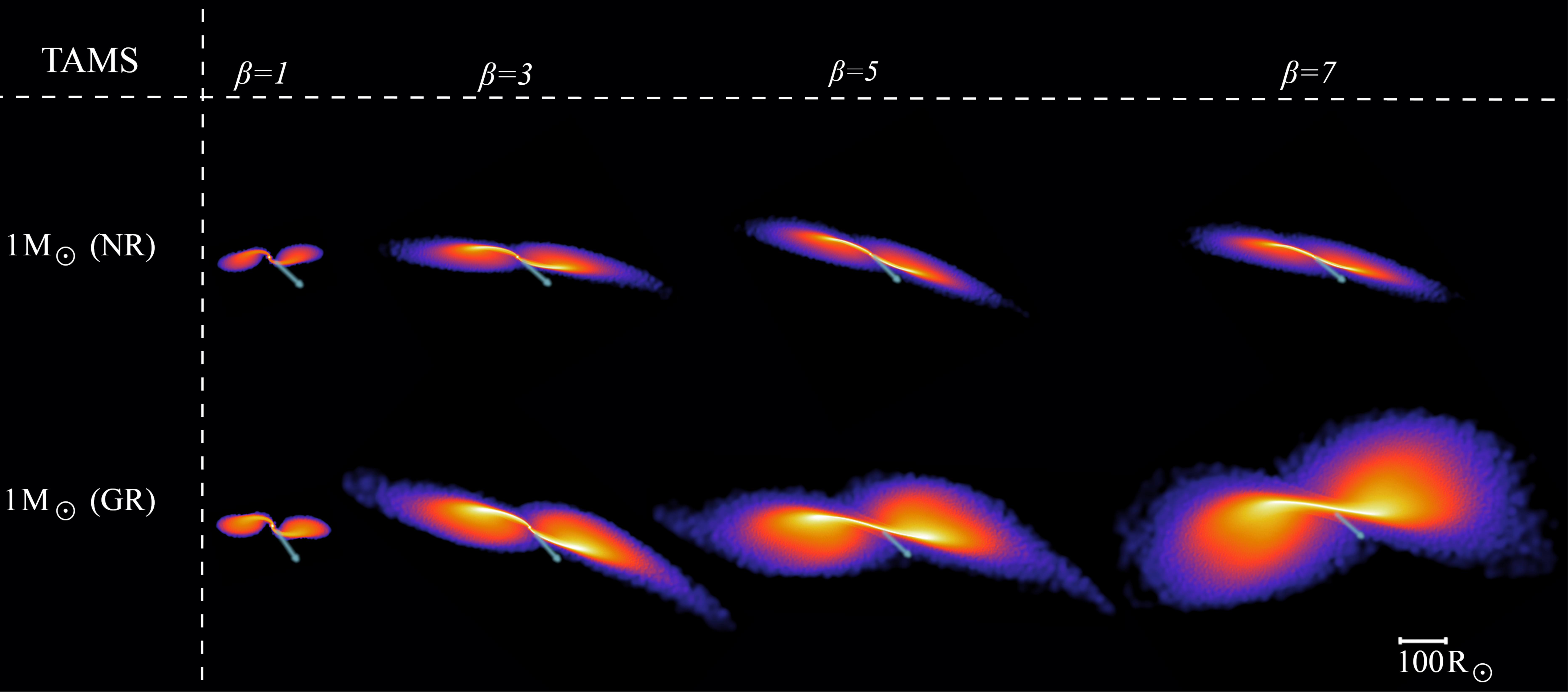}
		\end{minipage}
	
		\begin{minipage}[b]{0.903\linewidth}
		%\hspace*{0.0006\textwidth}
		\includegraphics[width=\textwidth]{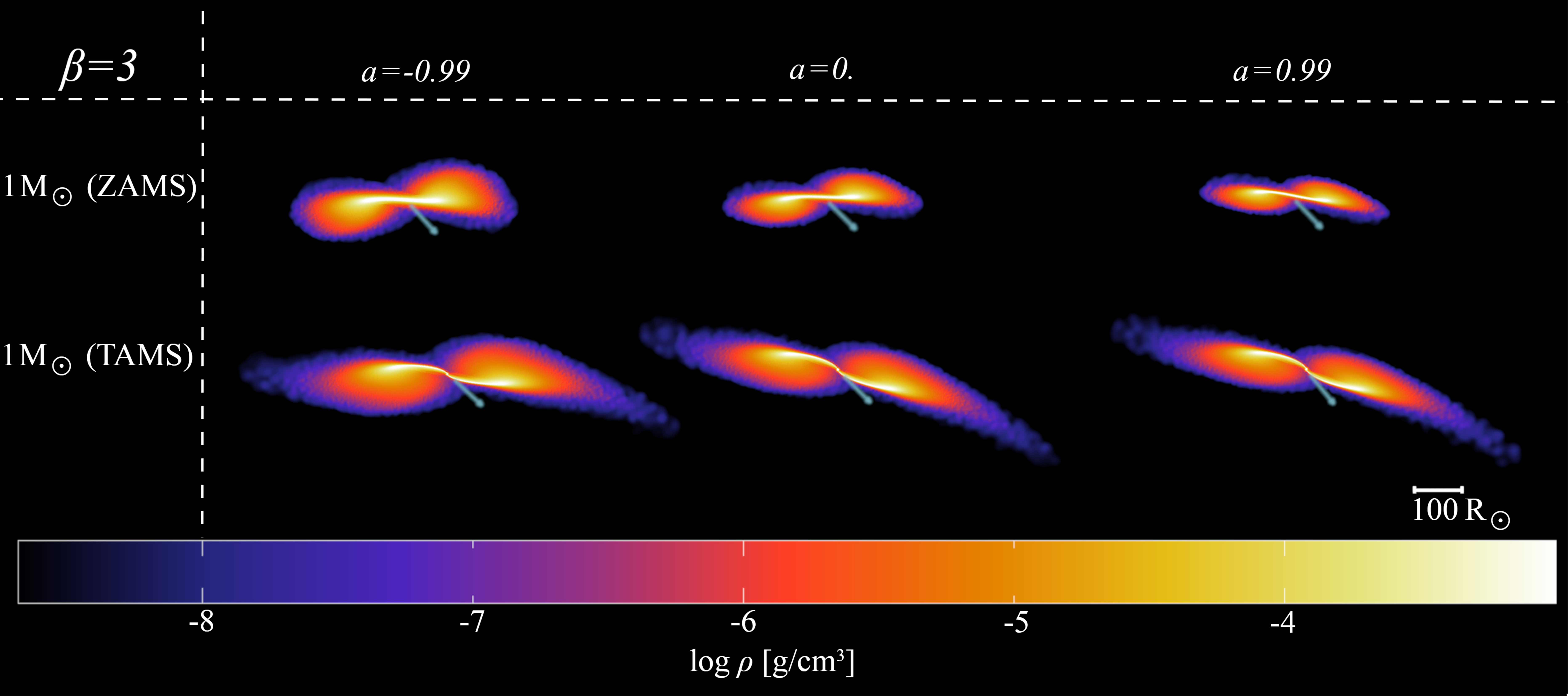}
		\end{minipage}
	
		\caption{Density slices of TDE simulations of $1\, \mathrm{M_\odot}$ stars at $\approx 15\, t_\mathrm{dyn}$ after the pericenter passage. Arrows indicate the direction to the SMBH. We rotate slices of the debris configurations so that the directions towards the SMBH are aligned. \emph{Top}: density slices of ZAMS stars after a disruption by a non-rotating SMBH with a Newtonian (first row) and a GR (second row) description of gravity. \emph{Middle}: density slices of TAMS stars after a disruption by a non-rotating SMBH with a Newtonian (first row) and a GR (second row) description of gravity. \emph{Bottom}: density slices of ZAMS (first row) and TAMS (second row) stars after a disruption by a rotating SMBH for $\beta=3$. }
		\label{s:4a}
	\end{figure*}

	Consequences of the effect of tidal field are also visible in the debris configurations before the second passage. Figure \ref{s:4a} shows density slices of the debris $\approx 15 t_\mathrm{dyn}$ after the disruption of $1\, \mathrm{M_\odot}$ ZAMS (top row) and TAMS (middle row) stars in a NR and a GR gravitational field for various $\beta$. $t_\mathrm{dyn}$ is the dynamical times scale $t_\mathrm{dyn} = \left ( \sqrt{G\overline{\rho}}  \right)^{-1}$. Disruptions in a relativistic gravitational field result in wider and more extended tidal streams in the orbital plane indicating that the debris is moving on a wider range of elliptical orbits. This also indicates that the disruptions are stronger in a GR than in a NR gravitational field for the same $\beta$. 
	
	Older stars are more centrally concentrated and have less bound envelopes. Consequently, disruptions of these stars lead to debris configurations with more torqued and elongated tidal streams. The exception is the disruption of a ZAMS star for $\beta=7$. In this case, the debris has been stretched almost to the maximum value due to the short orbital time scale and is on the verge of the second passage.
	
	Figure \ref{s:4a} (bottom row) also illustrates the effect of SMBH's rotation on the debris configuration in the case of $1\, \mathrm{M_\odot}$ stars for $\beta=3$. Disruptions of stars on retrograde orbits result in wider and more elongated tidal streams, similarly to stronger encounters. Furthermore, the effect of SMBH's rotation is more apparent in debris configurations of ZAMS stars. This is a consequence of shorter pericenter distances (for the same $\beta$) for younger stars --- the importance of relativistic effects and effects due to the SMBH's rotation increases for closer encounters.

	\subsection{Mass fallback rate of the debris}

\begin{figure*} [htb!]
		\centering
		\begin{minipage}[b]{.49\linewidth}
			\includegraphics[width=\textwidth]{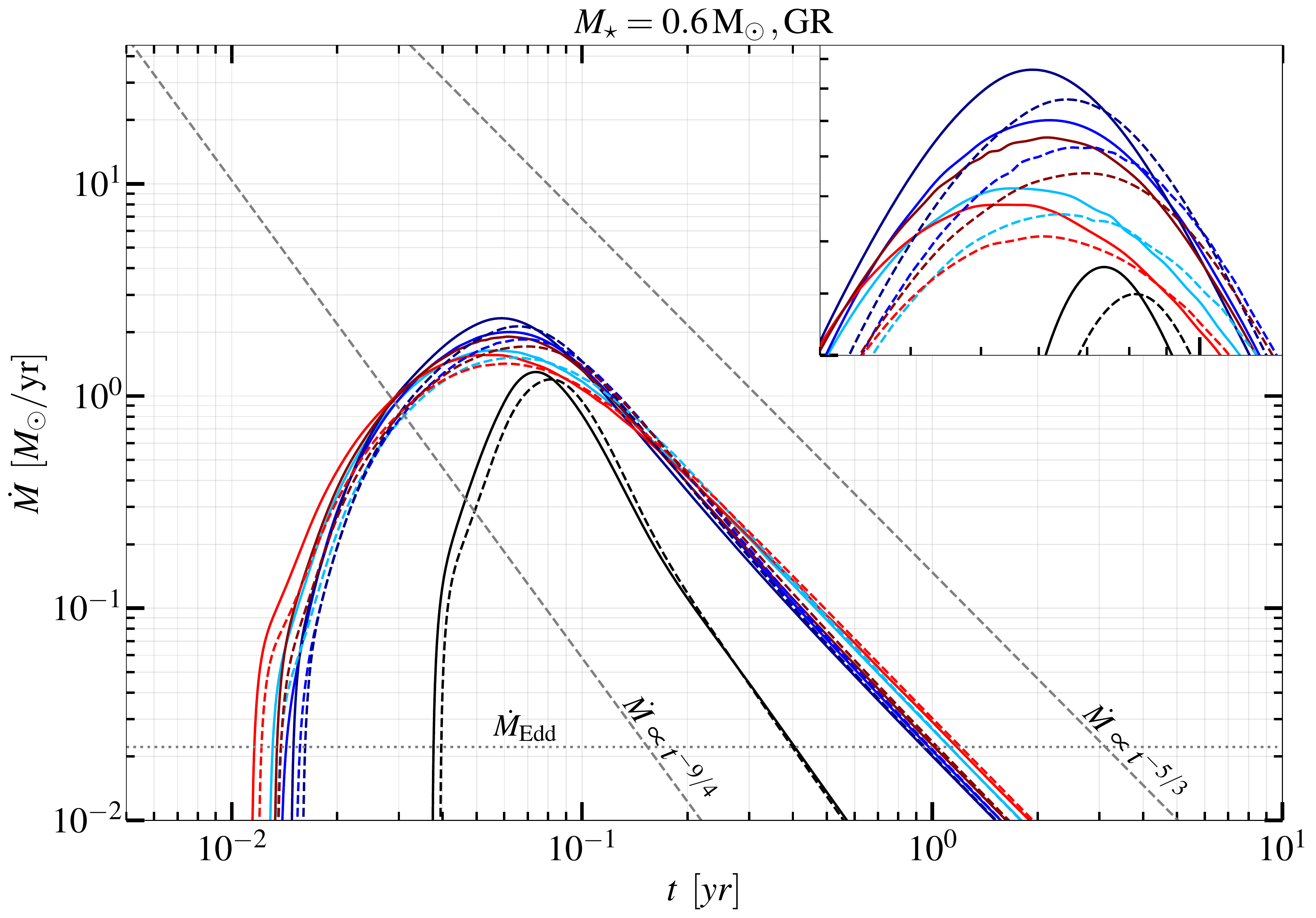}
		\end{minipage}\hfill
		\begin{minipage}[b]{.49\linewidth}
			\includegraphics[width=\textwidth]{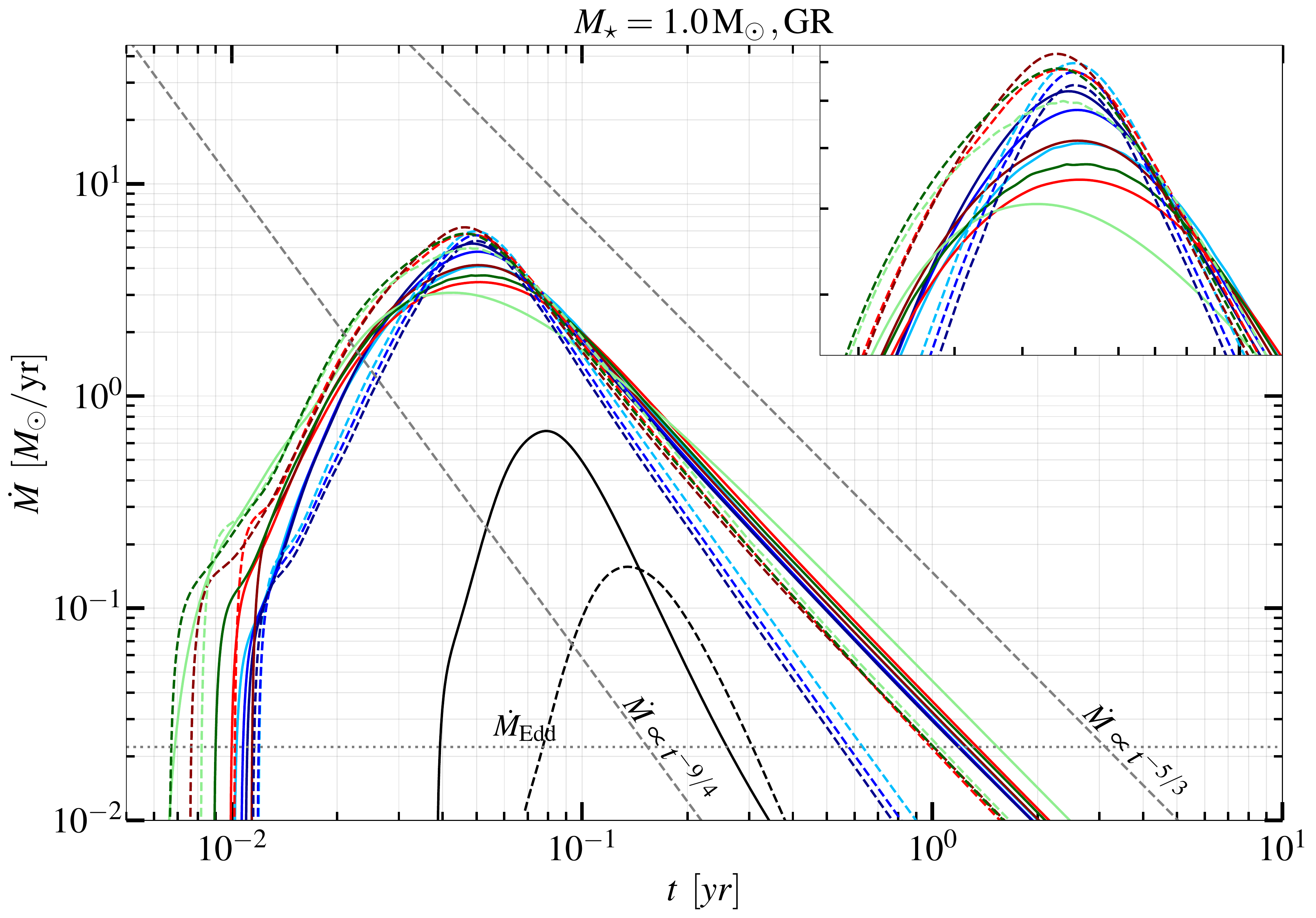}
		\end{minipage}
	
	\begin{minipage}[b]{.49\linewidth}
		\includegraphics[width=\textwidth]{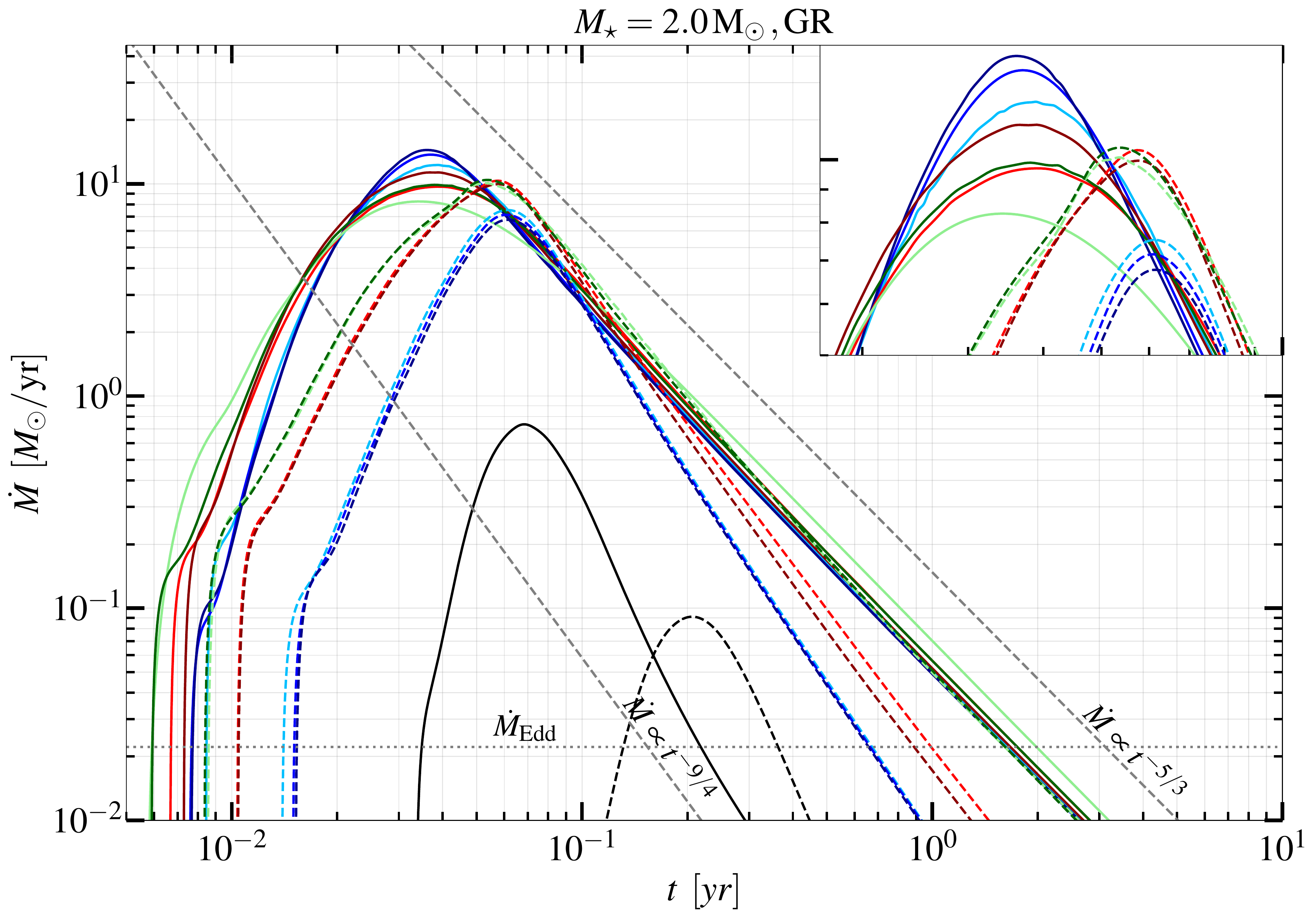}
	\end{minipage}\hfill
	\begin{minipage}[b]{.49\linewidth}
		\includegraphics[width=\textwidth]{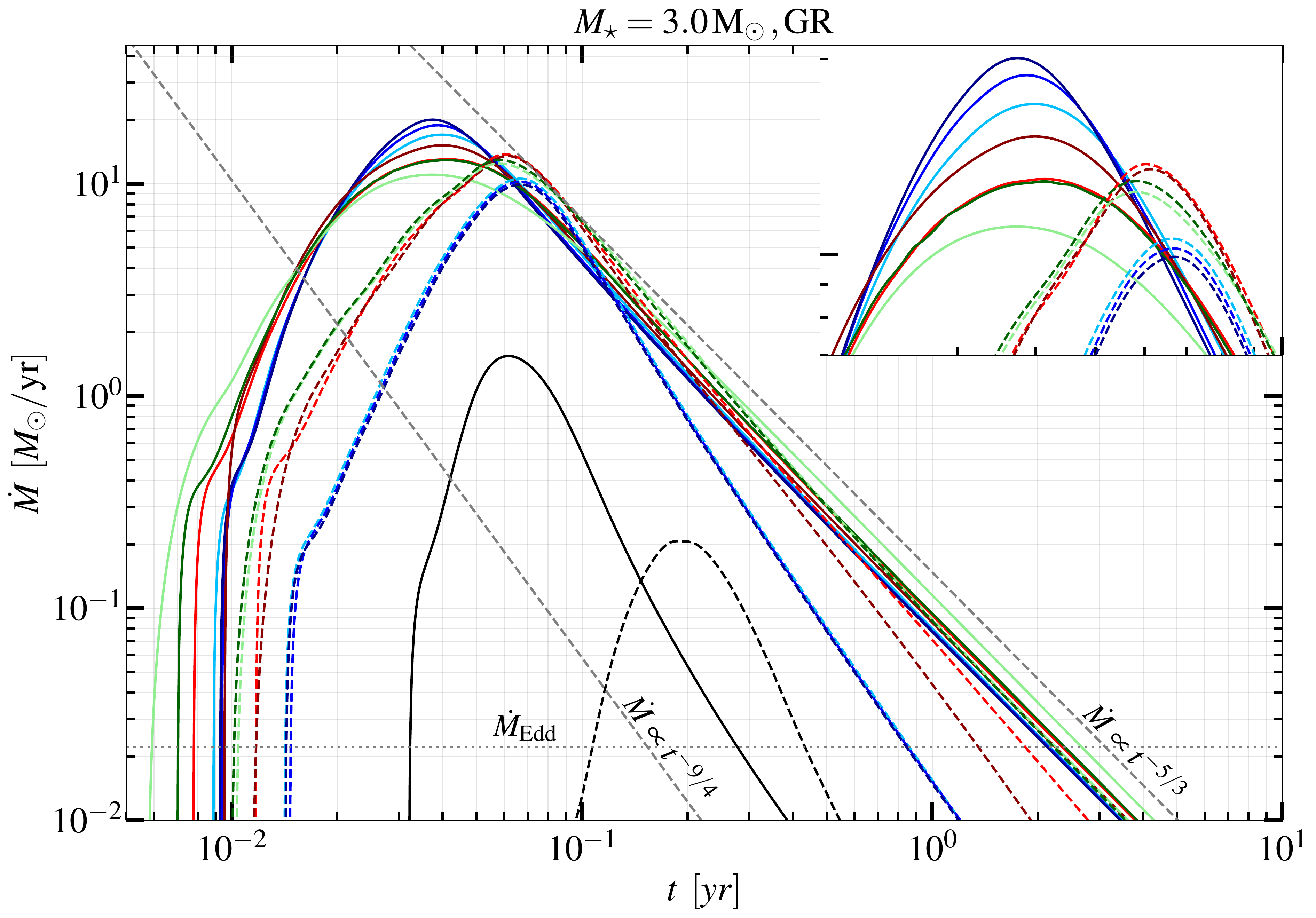}
	\end{minipage}

				\vspace*{0.1cm}
		\begin{minipage}[b]{\linewidth}
			\centering
			\includegraphics[width=0.45\textwidth]{dotM_legend_horizontal_crop.pdf}
		\end{minipage}
		\vspace*{-0.1cm}
		\caption{$\dot{M}$ for stars with different masses and ages. Horizontal dotted line indicates the Eddington accretion rate of a $10^6\, \mathrm{M_\odot}$ SMBH. Diagonal dotted lines represent power-law curves: $t^{-5/3}$ for total stellar disruptions, and $t^{-9/4}$ for partial stellar disruptions. Boxes in the top right corners of individual plots show a zoomed-in region near the peak of $\dot{M}$ curves.}
		\label{fig:dotM}
	\end{figure*}

	The resulting $\dot{M}$ curves are shown in Figure \ref{fig:dotM}. The characteristic properties of the curves ($\dot{M}_\mathrm{peak}$, $t_\mathrm{peak}$, $t_\mathrm{Edd}$, $t_{>0.5\dot{M}_\mathrm{peak}}$, $\tau_\mathrm{rise}$, $n_\infty$) differ significantly.

\begin{figure*} [htb!]
		\centering
	\begin{minipage}[b]{.49\linewidth}
	\includegraphics[width=\textwidth]{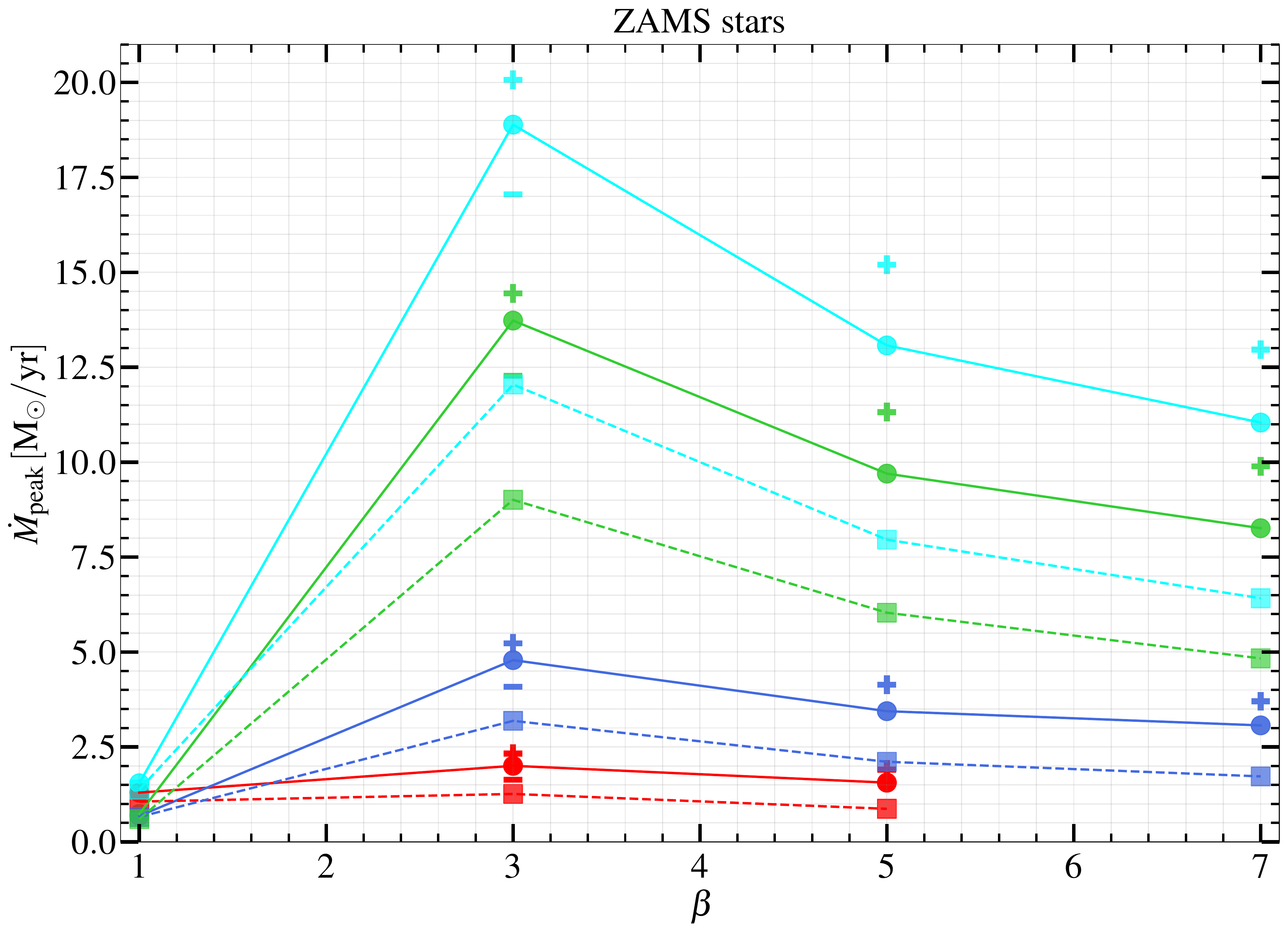}
\end{minipage}\hfill
\begin{minipage}[b]{.49\linewidth}
	\includegraphics[width=\textwidth]{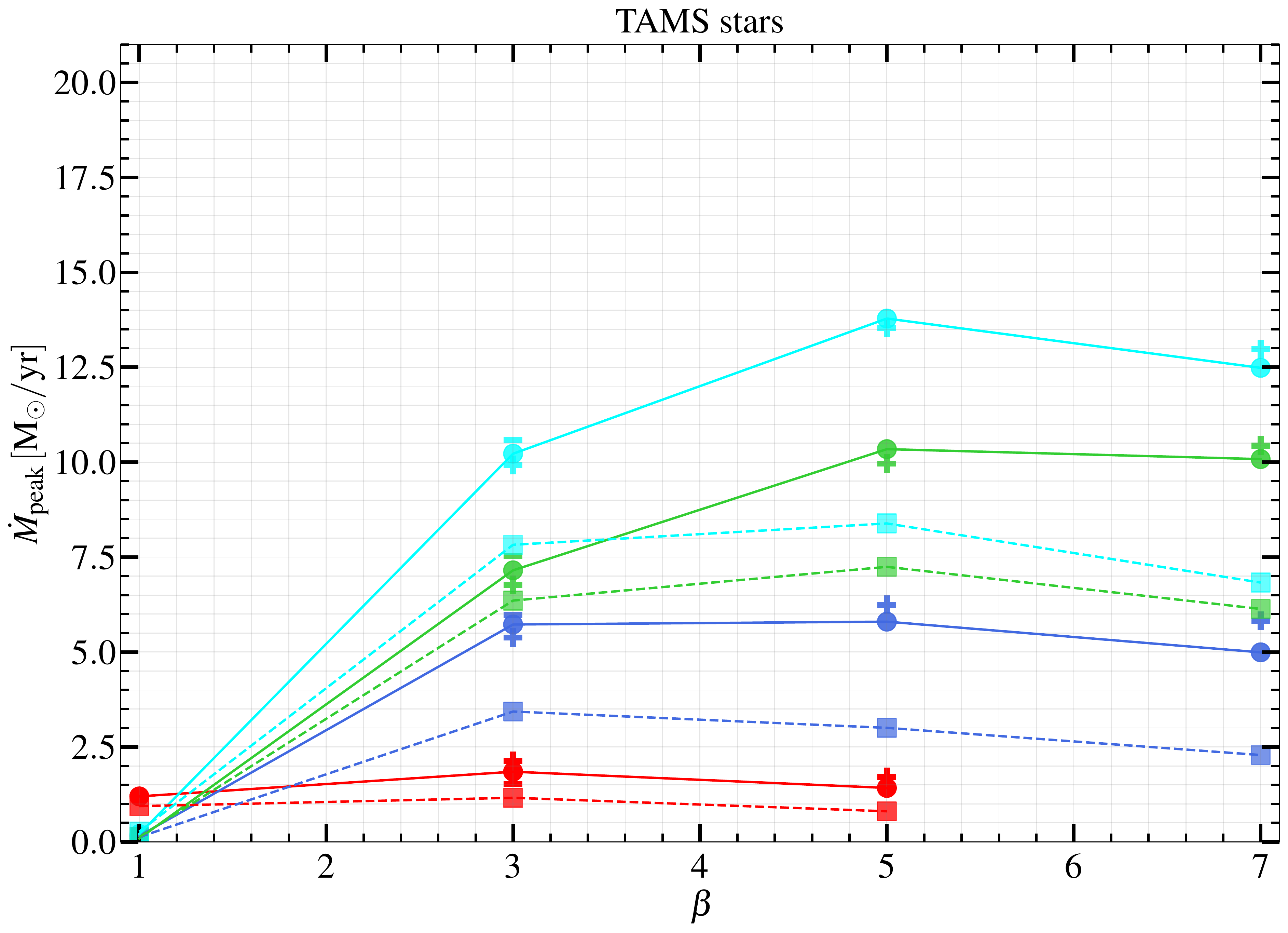}
\end{minipage}
\vspace*{0.1cm}
\begin{minipage}[b]{\linewidth}
			\centering
	\includegraphics[width=0.85\textwidth]{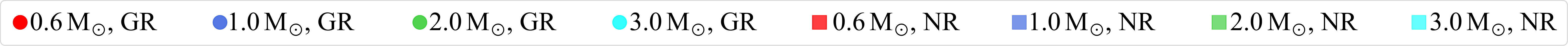}
\end{minipage}
\vspace*{-0.1cm}
		\caption{Peak values of $\dot{M}$ for ZAMS (left) and TAMS (right) stars for different $M_\star$, $a$ and $\beta$. "+" and "-" symbols indicate results from disruptions of stars on prograde and retrograde orbits, respectively. Results for GR and NR simulations are indicated with "$\bullet$" and "$\blacksquare$" symbols, respectively. }
		\label{s:5}
	\end{figure*}

	A detailed comparison of peak values $\dot{M}_\mathrm{peak}$ is seen in Figure \ref{s:5}. For NR encounters values of $\dot{M}_\mathrm{peak}$ are lower than for GR encounters with the same encounter parameters, which we contribute to a steeper gradient of the tidal field in GR. We note that the relative differences between NR and GR values of $\dot{M}_\mathrm{peak}$ increase with $\beta$ as the pericenter distance decreases and the general relativistic effects increase.

	From Equation (\ref{eq_dotMpeak}) it is expected that $\dot{M}_\mathrm{peak}$ increases with the stellar mass at a fixed mass loss. This behaviour can be seen in values of $\dot{M}_\mathrm{peak}$ for ZAMS stars at $\beta=3,5,7$ where TTDEs occur. We recover a similar dependence on the parameter $\beta$ as \citet{Law_Smith_2019} --- $\dot{M}_\mathrm{peak}$ increases up to approximately $\beta \approx \beta_\mathrm{crit}$, when a TTDE occurs, and then declines with a slower rate. 
	
	The effect of SMBH's rotation is seen in lower $\dot{M}_\mathrm{peak}$ values for total disruptions ($\Delta M/M_\star\approx1$) of stars on retrograde and higher for stars on prograde orbits. This effect is again similar to the dependency on the parameter $\beta$. In partial disruptions, the trend reverses: disruptions of stars on prograde orbits result in $\dot{M}$ curves with lower peak values because the amount of lost mass decreases with $a$. The effect of SMBH's rotation is greater for closer encounters --- for more massive stars and older stars at $\beta=3$ the peak values of the fallback rate are within $\lessapprox 2\%$, while for disruptions of ZAMS stars at $\beta=7$ encounters the differences can be $\approx 15\%$.

	\subsubsection{Characteristic time scales}
	
	We calculate three different characteristic times: time to the peak $t_\mathrm{peak}$, duration of the super-Eddington phase $t_\mathrm{Edd}$ (during which $\dot{M}>\dot{M}_\mathrm{Edd}$) and duration $t_\mathrm{>0.5\dot{M}_\mathrm{peak}}$ (during which $0.5\dot{M}_\mathrm{peak}<\dot{M}<=\dot{M}_\mathrm{peak}$).
	
\begin{figure*} [htb!]
		\centering
			\begin{minipage}[b]{.49\linewidth}
			\includegraphics[width=\textwidth]{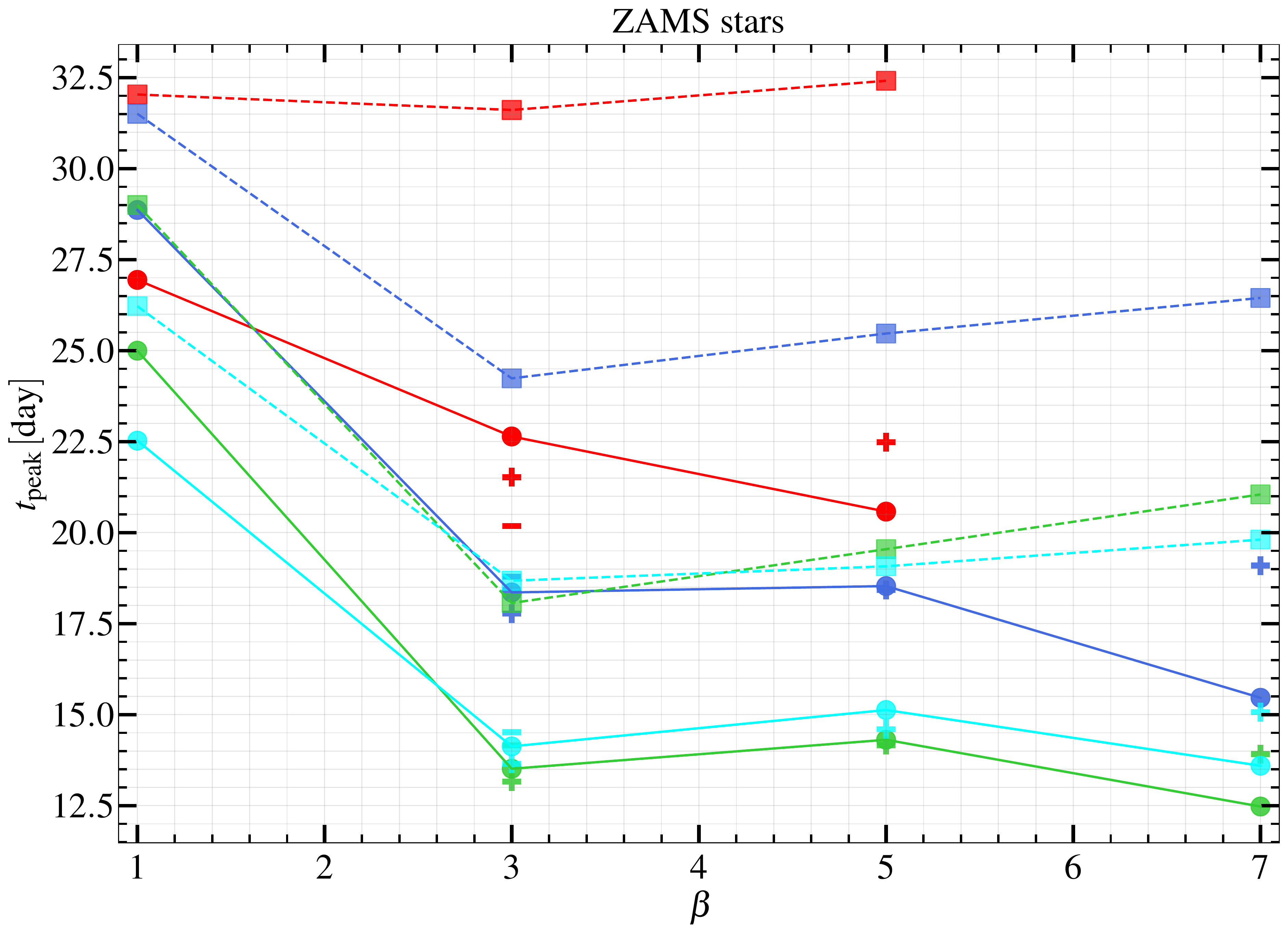}
		\end{minipage}\hfill
		\begin{minipage}[b]{.49\linewidth}
			\includegraphics[width=\textwidth]{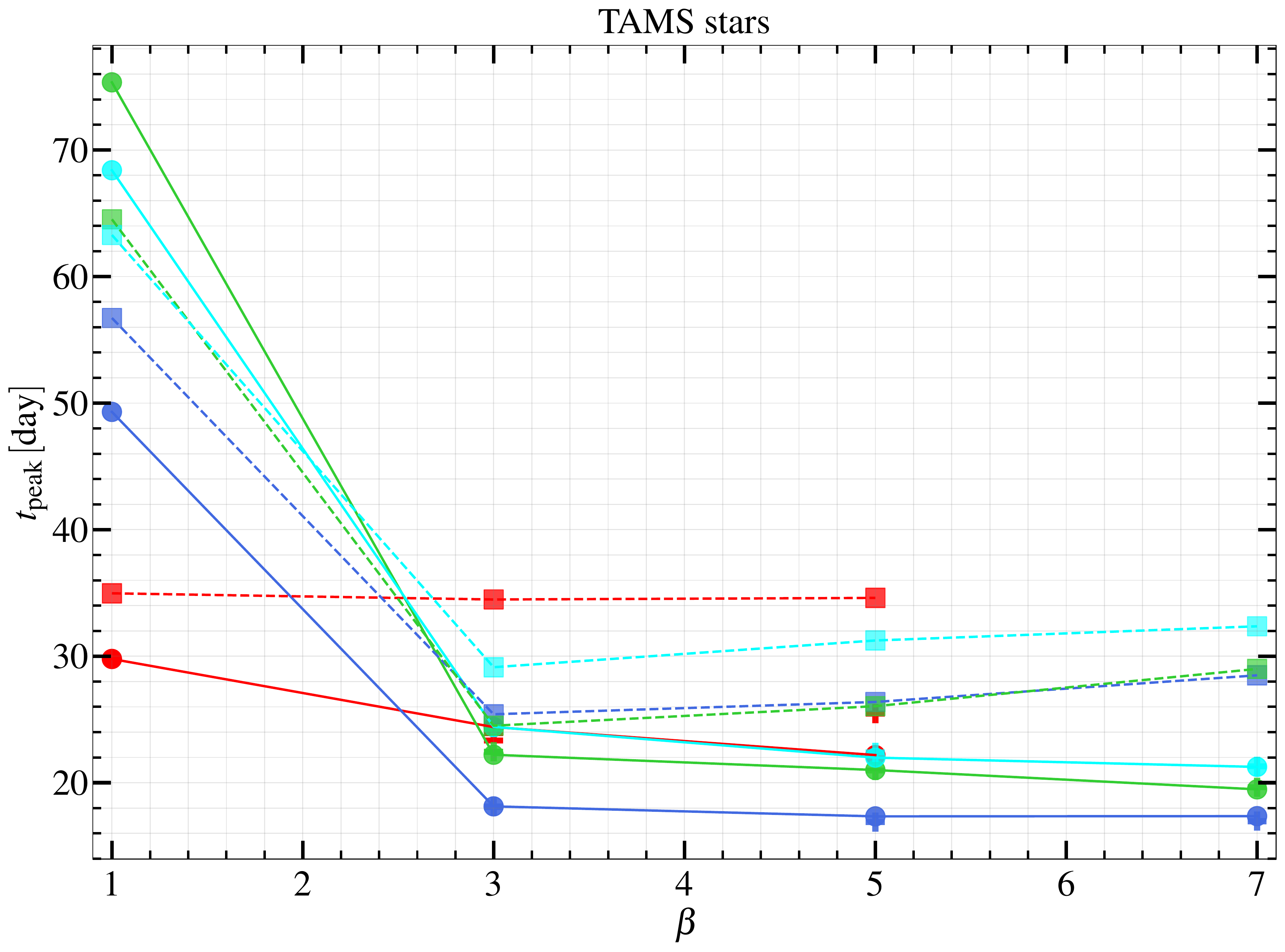}
		\end{minipage}
	
	\begin{minipage}[b]{.49\linewidth}
	\includegraphics[width=\textwidth]{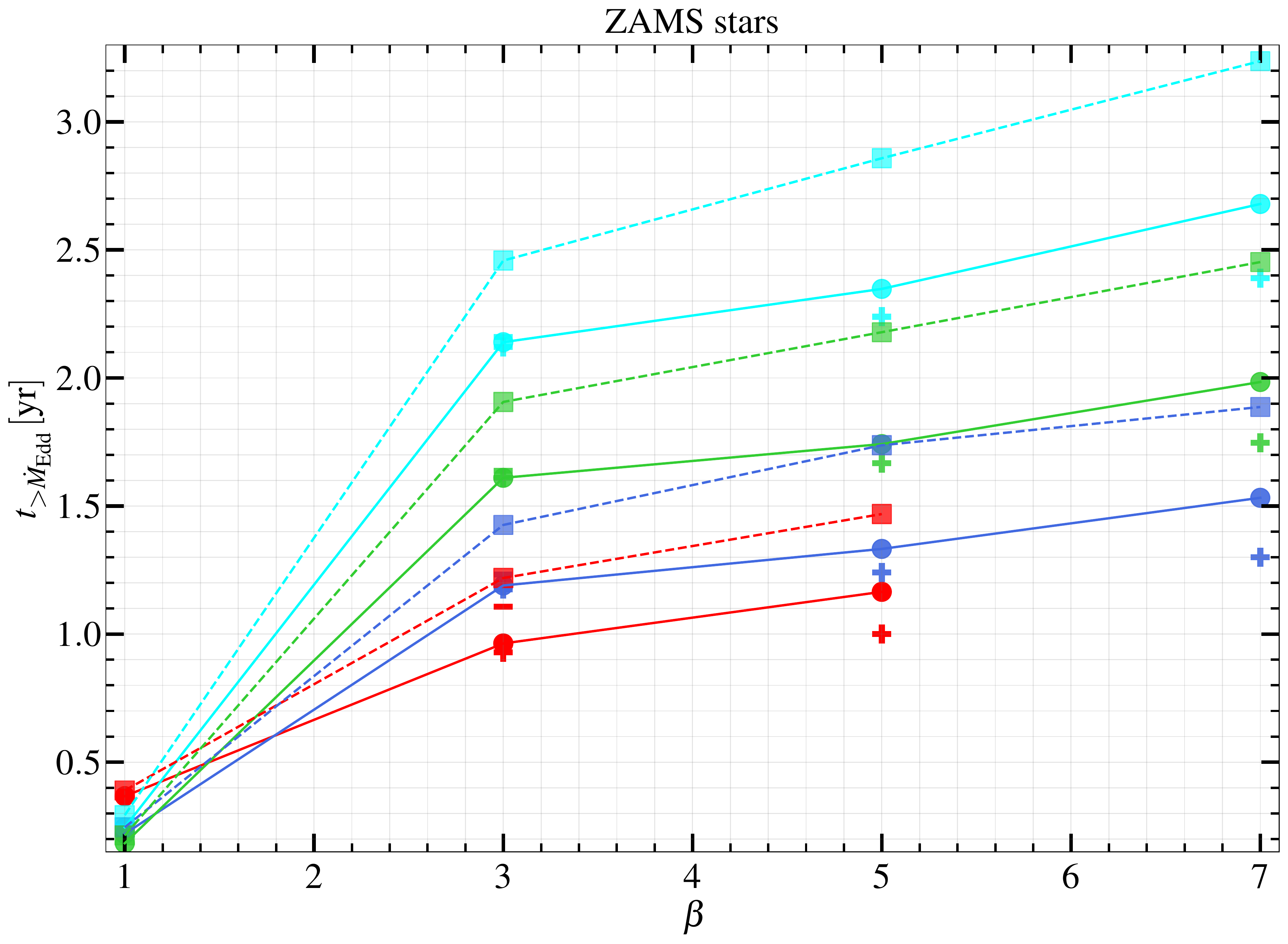}
\end{minipage}\hfill
\begin{minipage}[b]{.49\linewidth}
\includegraphics[width=\textwidth]{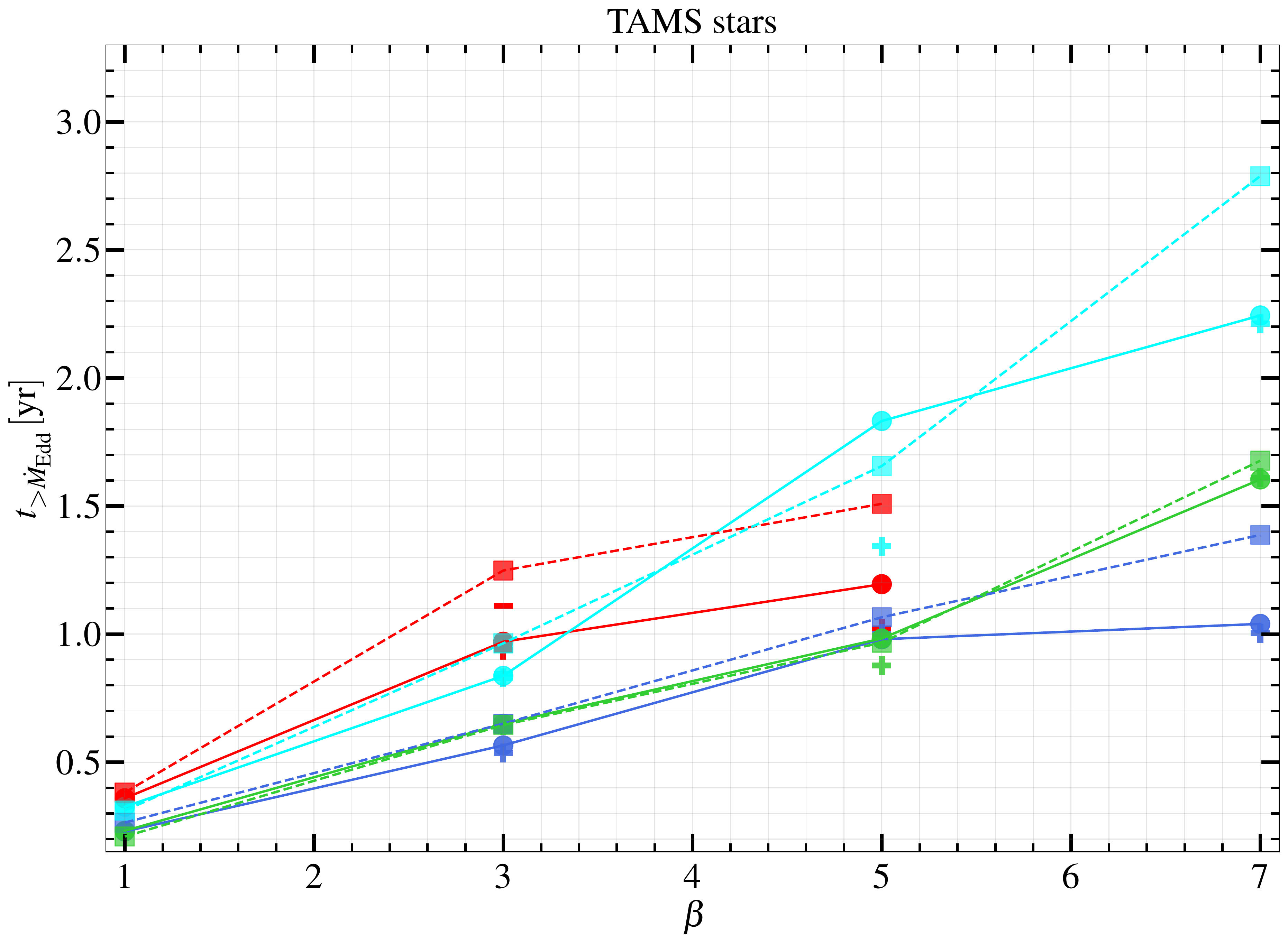}
\end{minipage}

\begin{minipage}[b]{.49\linewidth}
\includegraphics[width=\textwidth]{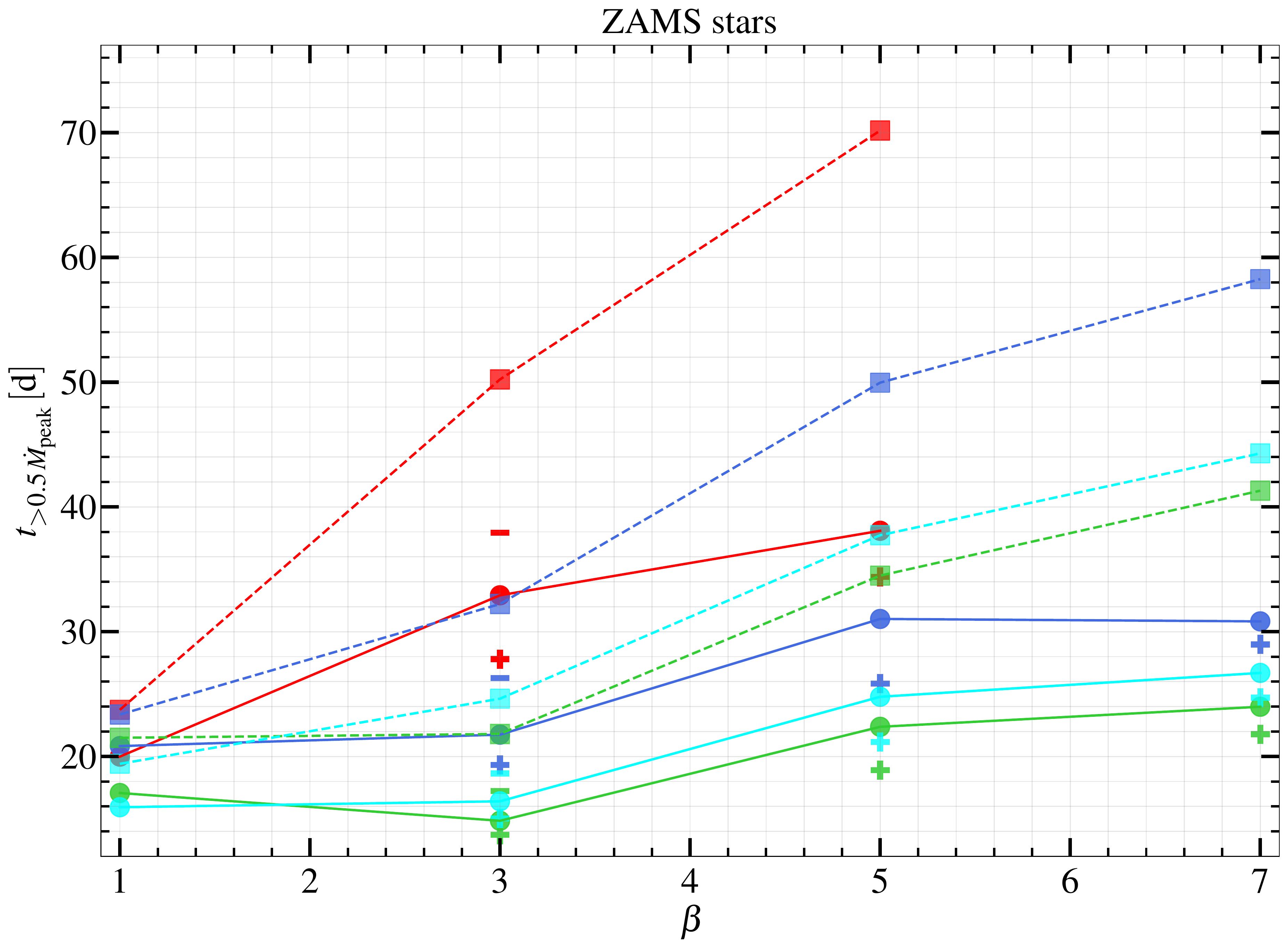}
\end{minipage}\hfill
\begin{minipage}[b]{.49\linewidth}
\includegraphics[width=\textwidth]{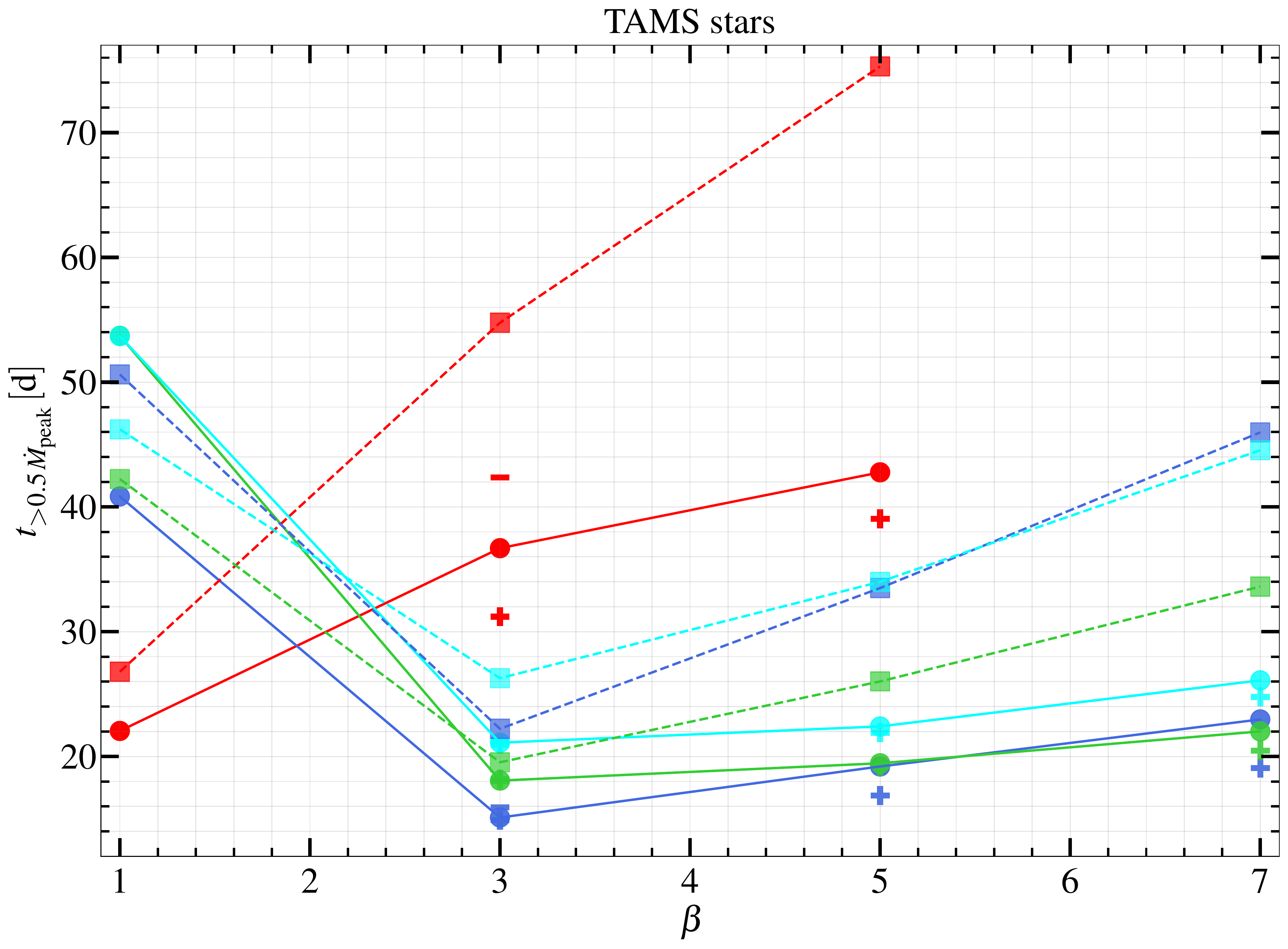}
\end{minipage}

		\vspace*{0.1cm}
		\begin{minipage}[b]{\linewidth}
			\centering
			\includegraphics[width=0.85\textwidth]{legend_horizontal_char_times.pdf}
		\end{minipage}
		\vspace*{-0.1cm}
		
		\caption{
			Characteristic time scales $t_\mathrm{peak}$ (\emph{top}), $t_\mathrm{Edd}$ (\emph{middle}) and  $t_\mathrm{>0.5\dot{M}_\mathrm{peak}}$ (\emph{bottom}) for ZAMS (left) and TAMS (right) stars for different $M_\star$, $a$ and $\beta$. "+" and "-" symbols indicate results from disruptions of stars on prograde and retrograde orbits, respectively. Results for GR and NR simulations are indicated with "$\bullet$" and "$\blacksquare$" symbols, respectively.}
		\label{s:10}
	\end{figure*}

	In Figure \ref{s:10} we see the results for $t_\mathrm{peak}$ (top row), $t_\mathrm{Edd}$ (middle row) and $t_\mathrm{>0.5\dot{M}_\mathrm{peak}}$ (bottom row). Due to a stronger SMBH's tidal field in a relativistic description of gravity there is more material on less energetic orbits (see Figure \ref{fig:gr_vs_nr}). GR TDEs produce fallback rate curves that reach peak values sooner, with a shorter duration of both $t_\mathrm{Edd}$ and $t_\mathrm{>0.5\dot{M}_\mathrm{peak}}$. 
	
	$t_\mathrm{peak}$ decreases with $\beta$, however, for $\beta \gtrapprox \beta_\mathrm{crit}$ it decreases at a slower rate. General trend of $t_\mathrm{Edd}$ and $t_\mathrm{>0.5\dot{M}_\mathrm{peak}}$ is that both quantities increase with the strength of disruption. In GR encounters of ZAMS stars $t_\mathrm{Edd}$ changes most drastically during the transition between PTDEs and TTDEs (by a factor of $\approx 2-4$) and continues to steadily increase with $\beta$. GR values of $t_\mathrm{>0.5\dot{M}_\mathrm{peak}}$ change by a maximum of factor $\approx 1.5$. $t_\mathrm{Edd}$ and $t_\mathrm{>0.5\dot{M}_\mathrm{peak}}$ of TAMS stars exhibit a similar pattern to ZAMS stars, with the exception being $t_\mathrm{>0.5\dot{M}_\mathrm{peak}}$ for $\beta=1$. In this case durations of $t_\mathrm{>0.5\dot{M}_\mathrm{peak}}$ are substantially larger than for $\beta=3$. We contribute this to larger pericenter distances during the disruptions of TAMS stars. As a consequence a larger amount of the debris is moving on more energetic orbits and $\mathrm{d}M/\mathrm{d}\epsilon$ curves are shifted towards more positive $\epsilon$. This effect is only visible for disruption of TAMS stars on $\beta=1$ orbits, where the pericenter distances can differ up to a factor of two from the pericenter distance of their ZAMS counterparts at $\beta=1$. For larger $\beta$ the mass over energy distributions span a more similar range of $\epsilon$ and consequently also more similar range of return times of the debris (see Figure \ref{s:4}).
	
	At a fixed $\beta$ less massive stars have shorter pericenter distances and experience a stronger tidal field. Therefore, the debris spans a wider range of elliptic orbits. However, more massive stars can have higher integrated values of $\dot{M}$, depending on the amount of lost mass. In the case of TTDEs of ZAMS stars this effect is seen in longer $t_\mathrm{Edd}$ for more massive stars, which is also supported by Equation (\ref{eq_tedd}). On the other hand, $t_\mathrm{peak}$ decreases with the stellar mass, contrary to what one might expect from Equation (\ref{eq_tpeak}). For TAMS stars, which have less bound envelopes and are mostly PTDEs, the interpretation of results is more intricate. For instance, at $\beta=5$ a $1\, \mathrm{M_\odot}$ star gets closer to the SMBH than $2\, \mathrm{M_\odot}$ and $3\, \mathrm{M_\odot}$ stars and has a lower value of mass lost. As a consequence, it has a comparable $t_\mathrm{Edd}$ to a $2\, \mathrm{M_\odot}$ star. However, a $3\, \mathrm{M_\odot}$ star has more mass and a sufficiently larger (less bound) envelope, compensating for a weaker disruption due to a larger pericenter distance during the first passage, which results in a longer $t_\mathrm{Edd}$ than in the case of a $1\, \mathrm{M_\odot}$ star. A similar dependency on the stellar mass is observed in values of $t_\mathrm{>0.5\dot{M}_\mathrm{peak}}$ for TAMS stars. Disruptions of younger stars are mostly TTDEs and therefore result in higher values of $t_\mathrm{Edd}$ than for disruptions of older TAMS stars.

	In disruptions of ZAMS stars SMBH's rotation induces shorter durations of $t_\mathrm{Edd}$ and $t_\mathrm{>0.5\dot{M}_\mathrm{peak}}$ for prograde and longer for retrograde stellar orbits. The effect of SMBH's rotation is less apparent in trends of $t_\mathrm{peak}$ for low $\beta$ encounters. In high $\beta$ encounters  $t_\mathrm{peak}$ is longer for prograde orbits. Disruptions of TAMS stars by a rotating SMBH exhibit a similar trend to their ZAMS counterparts. The difference is that the results for rotating and non-rotating SMBH differ by a lower amount, due to a larger pericenter distance and consequently a lower effect of SMBH's rotation.

	\newpage
	\subsubsection{Early rise-time and late-time decline}

\begin{figure*} [htb!]
		\centering
						\begin{minipage}[b]{.49\linewidth}
			\includegraphics[width=\textwidth]{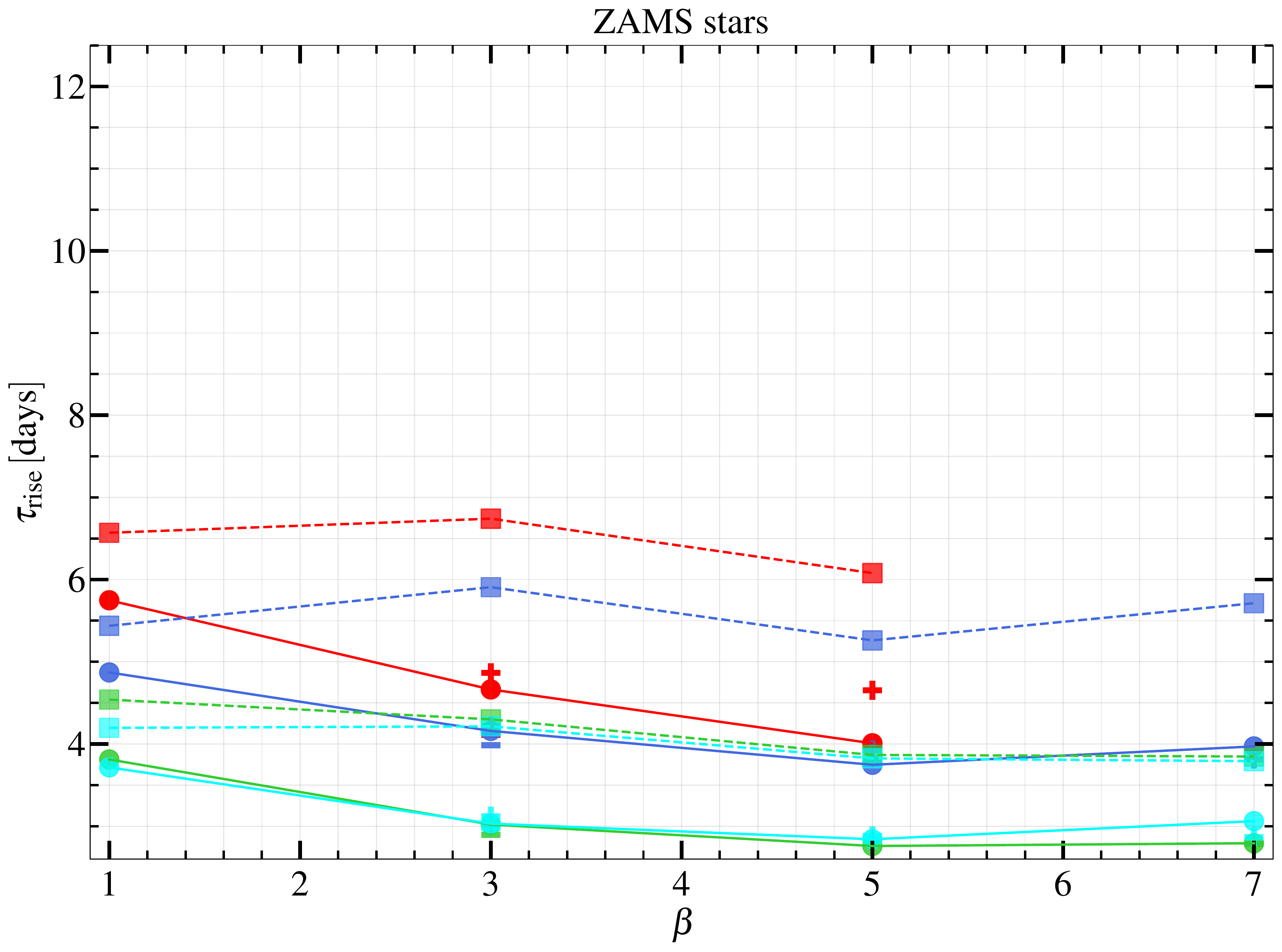}
		\end{minipage}\hfill
		\begin{minipage}[b]{.49\linewidth}
			\includegraphics[width=\textwidth]{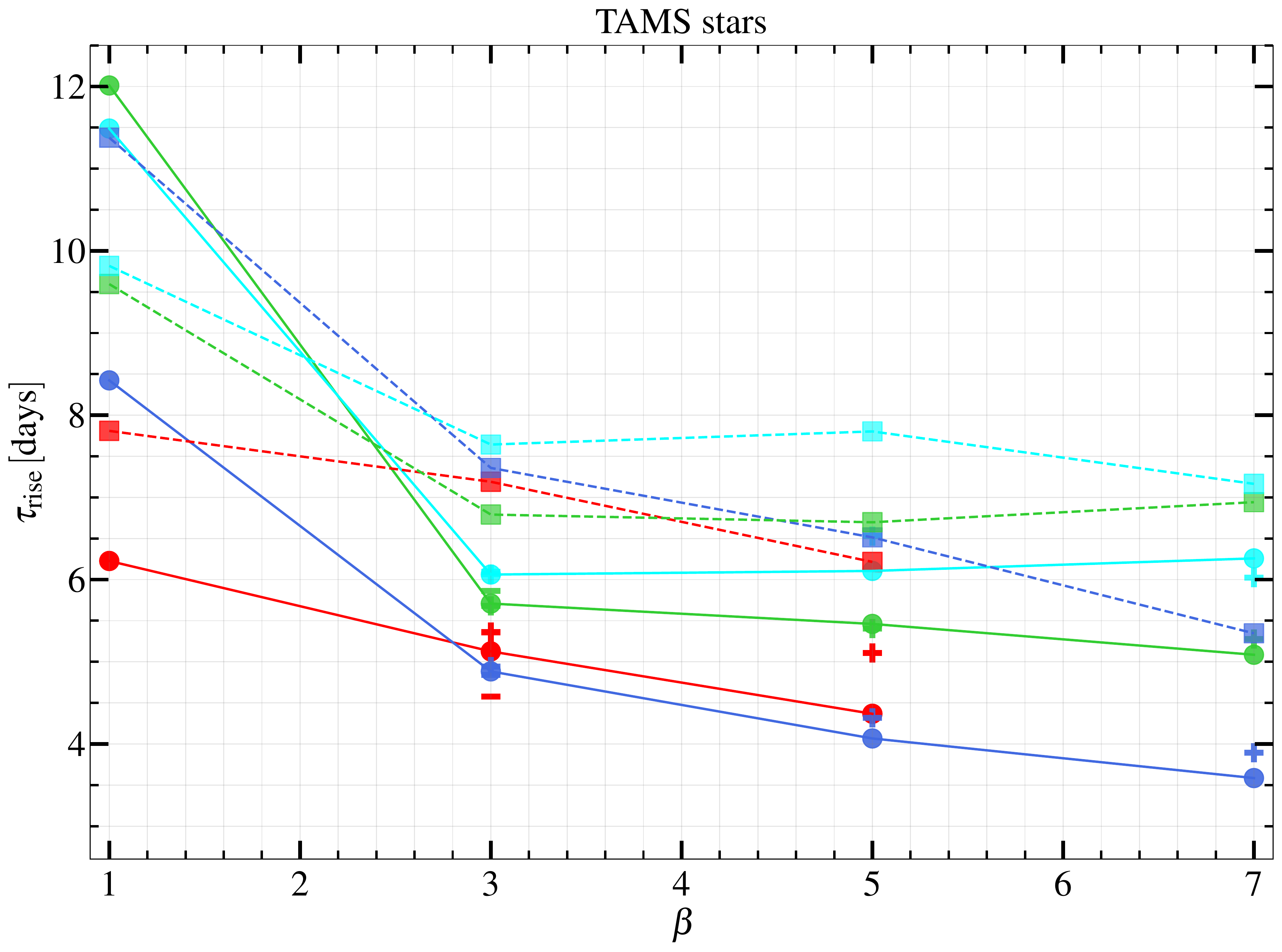}
		\end{minipage}
							\begin{minipage}[b]{.49\linewidth}
			\includegraphics[width=\textwidth]{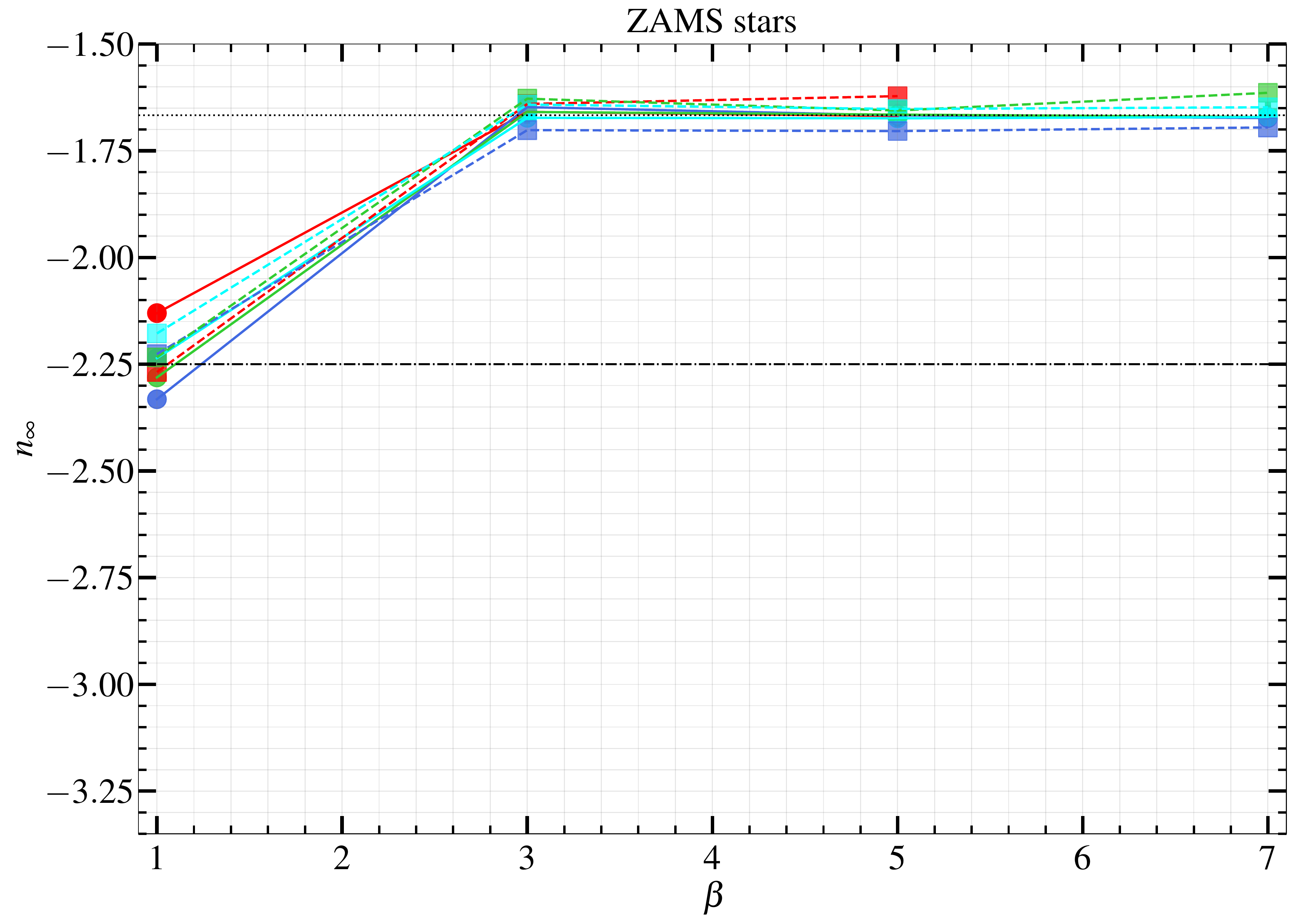}
		\end{minipage}\hfill
		\begin{minipage}[b]{.49\linewidth}
			\includegraphics[width=\textwidth]{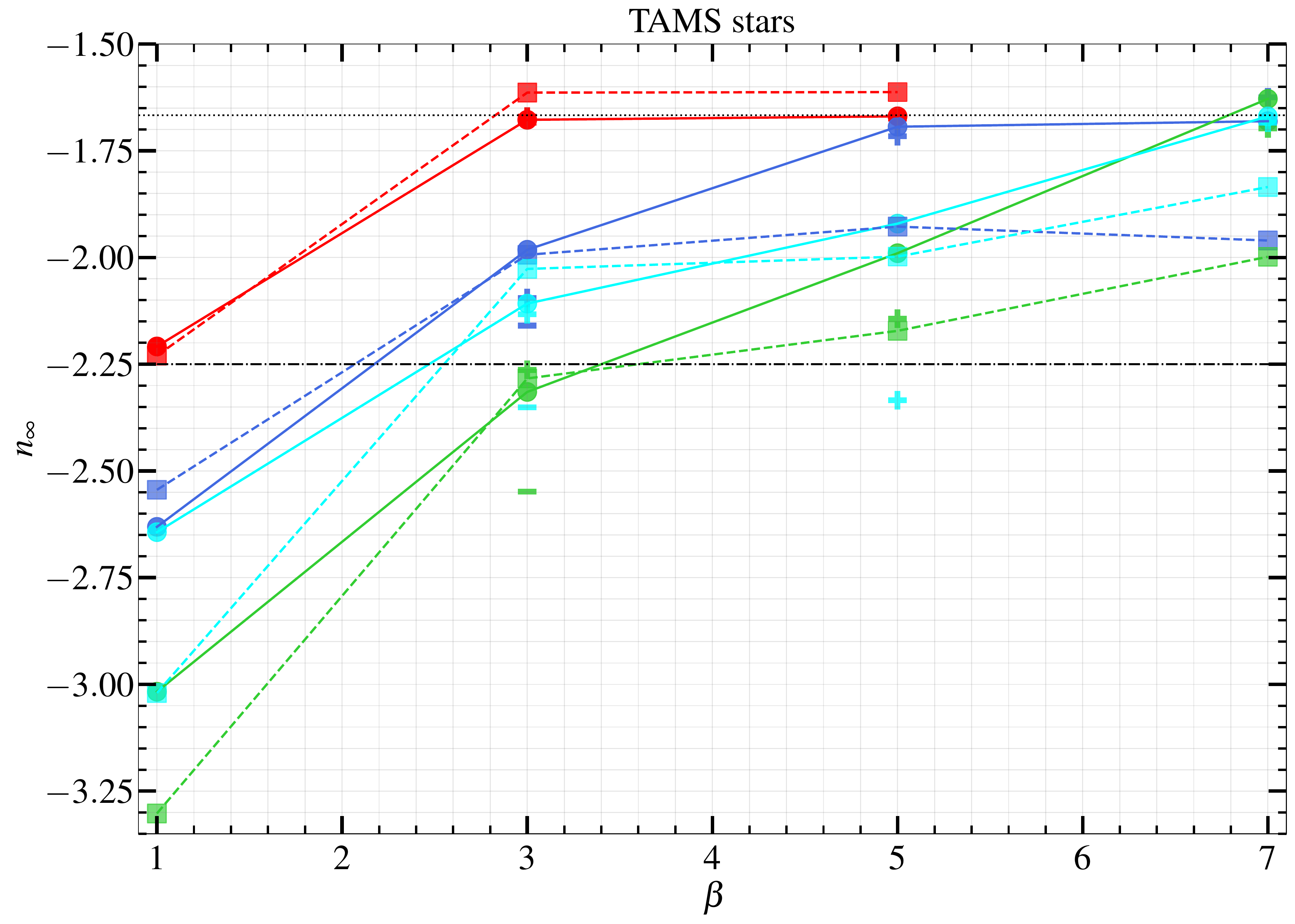}
		\end{minipage}

\vspace*{0.1cm}
\begin{minipage}[b]{\linewidth}
	\centering
	\includegraphics[width=0.85\textwidth]{legend_horizontal_char_times.pdf}
\end{minipage}
\vspace*{-0.1cm}

		\caption{
			Early rise-time $\tau_\mathrm{rise}$ (\emph{top}) and late-time slope $n_\infty$ (\emph{bottom})  for ZAMS (left) and TAMS (right) stars for different $M_\star$, $a$ and $\beta$. "+" and "-" symbols indicate results from disruptions of stars on prograde and retrograde orbits, respectively. Results for GR and NR simulations are indicated with "$\bullet$" and "$\blacksquare$" symbols, respectively. For partial disruption the analytically predicted slope is $n_\infty \approx -9/4$ and for total disruption it is $n_\infty \approx -5/3$, which are represented with a dashed and a dash-dotted line, respectively.
		}
		\label{s:11}
	\end{figure*}

	We also calculate early characteristic rise-time $\tau_\mathrm{rise}$ and late-time slope $n_\infty$. Obtained values are shown in Figure \ref{s:11}. Characteristic rise-time $\tau_\mathrm{rise}$ is defined as the standard deviation of a Gaussian function fitted to the early part of $\dot{M}$ curves, between $20-80\%$ of $\dot{M}_\mathrm{peak}$, and is analogous to the $e$-folding time. Therefore, lower values of $\tau_\mathrm{rise}$ indicate a steeper rise of the fallback rate curves. Disruptions of stars in a relativistic tidal field lead to higher peak values of $\dot{M}$ and shorter times to the peak. In agreement with this, we find that relativistic disruptions result in $\dot{M}$ curves with lower values of $\tau_\mathrm{rise}$. An exception is the $2\, \mathrm{M_\odot}$ TAMS star at $\beta=1$, which has a substantially longer $t_\mathrm{peak}$ and therefore longer $\tau_\mathrm{rise}$.
	
	As the strength of encounters increases, $\tau_\mathrm{rise}$ becomes shorter. Furthermore, we find that disruptions of more centrally concentrated ZAMS stars (at a fixed $\beta$) lead to steeper rises of $\dot{M}$. In the case of partial disruptions of TAMS stars we find a different trend. TAMS stars have a lower density in the outer layers and these layers are more prone to disruption. As a consequence the debris is further stretched out and the density gradient in tidal tails is lower, which leads to a shallower rise slope. Furthermore, we find that the rotation of the black hole increases $\tau_\mathrm{rise}$ for prograde orbits and decreases it for retrograde orbits. 
	
	For total disruptions $n_\infty$ approaches $-5/3$, in agreement with previous numerical studies of the $\dot{M}$ and theoretical predictions \citep{rees, Guillochon_2013, lawsmith2020stellar}. Theoretical analysis for partial disruptions predicts  $n_\infty \approx -9/4$ \citep{Coughlin_2019}. We find that for PTDEs there is a much larger scatter present around the predicted value. The SMBH's rotation can affect the late-time slope by both decreasing or increasing its value for the same $\beta$. However, we find no particular trend.

	\section{Discussion} \label{sec:4}
	
	\subsection{Disruptions in relativistic and Newtonian tidal fields}
	\citet{Kesden_2012} and \citet{Servin_2017} study the differences between the energy spread in GR and in NR and give two possible explanations. \citet{Kesden_2012} argues that, for the same $\beta$, GR disruptions would be stronger and the energy spread would be larger. This would result in earlier peak times, higher $\dot{M}_\mathrm{peak}$ and more narrow $\dot{M}$ curves. However, \citet{Servin_2017} argue that in GR a star would be disrupted further away from a SMBH, due to the steeper potential well. This would result in a smaller energy spread in a relativistic tidal field and the effect on the $\dot{M}$ would be the opposite --- later peak times, lower $\dot{M}_\mathrm{peak}$ and broader $\dot{M}$ curves.
	
	We determine the relativistic effects by comparing simulations of stellar disruptions at the same pericenter distances. In agreement with \citet{Kesden_2012} we find that disruptions in GR are stronger and produce a wider energy spread. The mass fallback rate of the debris reaches the peak value earlier and has a higher peak value. Furthermore, $\dot{M}$ curves in GR have a narrower shape than in NR --- values of $t_\mathrm{Edd}$ and $t_\mathrm{>0.5\dot{M}_\mathrm{peak}}$ are lower in a relativistic tidal field. We find that the rise-time of relativistic $\dot{M}$ is shorter.

	Relativistic effects on the energy spread and $\dot{M}$ were also studied by \citet{Cheng_2014} and \citet{Gafton_2019}. For a relativistic disruption of a $1\,\mathrm{M_\odot}$ polytropic star (for $\beta=1$) by a $10^6\, \mathrm{M_\odot}$ SMBH \citet{Cheng_2014} find that $\dot{M}_\mathrm{peak}$ is higher by $\approx 2\%$ (we find $\approx 4\%$ for a ZAMS star), while $t_\mathrm{peak}$ and $t_\mathrm{Edd}$ are longer by $\approx 10\%$  and $\approx 7\%$ , respectively, in comparison to NR encounters (we find that $t_\mathrm{peak}$ and $t_\mathrm{Edd}$ are shorter by $\approx 9\%$ and $\approx 10\%$, respectively, for a $1\,\mathrm{M_\odot}$ ZAMS star). Similar trend was also found by \citet{Gafton_2019}. For comparison purposes we also simulate a NR and a GR disruption of a $1\,\mathrm{M_\odot}$ polytropic star for $\beta=1$, and find the same trend as in disruptions of MESA stars. Therefore, we speculate that the discrepancies between our results and those by \citet{Cheng_2014} and \citet{Gafton_2019} are not due to different stellar density profiles, but are caused by a different implementation of the self-gravity and/or the usage of a different TDE simulation software. \citet{Gafton_2019} used an SPH code with a pseudo-relativistic self-gravity description, \citet{Cheng_2014} used a numerical method that computes debris properties in a Fermi-normal coordinate system with the Newtonian self-gravity, while we use an SPH code with a Newtonian self-gravity. We discuss this further in Section \ref{sec:caveats}.
	%However, we emphasize that currently there is no viable method that would test the validity of the self-gravity implementation. The most accurate method would be to calculate self-gravity by solving Einstein equations for every particles. However, this approach would severely increase the computational time.
	
	\citet{Servin_2017} and \citet{Stone_2019} propose another approach to compare the differences between disruptions in GR and NR gravitational potentials. This is done by considering disruptions of stars with the same values of the specific angular momentum magnitude $L$. In this case it is expected that stars with the same angular momentum magnitude would experience stronger tides  in GR at the pericenter. In our case, NR simulations have lower values of $L$ than GR simulations at a fixed $\beta$. However, it is possible to compare GR and NR disruptions with a similar value of $L$ by considering stellar orbits with a different $\beta$. For instance, a ZAMS star on an orbit with $\beta=7$ in a relativistic gravitational field has a value of $L$ similar (within $\approx 1\%$) to the one of a ZAMS star on an orbit with $\beta=5$ in a Newtonian gravitational field. Therefore, by taking $L$ as a proxy value for strength of the encounter (instead of the pericenter distance) we can compare the characteristics of $\dot{M}$ curves. For disruptions with the same value of $L$ the relative differences between GR and NR values of $\dot{M}$ characteristics are lower by a factor of $\approx 2$ (in comparison to relative differences in disruptions at the same pericenter distance), while the trend remains the same. For instance, the relative difference in $\dot{M}_\mathrm{peak}$ is $\approx 43\%$ (instead of $\approx 76\%$), while the difference in $t_\mathrm{Edd}$ is $\approx 13\%$ (instead of $\approx 27\%$).

	\subsection{The effect of the SMBH's rotation}
	SMBH's rotation influences the initial stellar orbit by affecting the pericenter distance: pericenter distances of stars on retrograde orbits are decreased, while pericenter distances of stars on prograde orbits are increased. Therefore, stars on prograde orbits experience a smaller tidal field, which induces a smaller spread in energy of the debris. Consequently, $\dot{M}$ curves of fully disrupted stars on prograde orbits have lower values of $\Delta M$, higher $\dot{M}_\mathrm{peak}$, longer $t_\mathrm{peak}$, are narrower (shorter durations of  $t_\mathrm{Edd}$ and $t_\mathrm{>0.5\dot{M}_\mathrm{peak}}$) and have longer $\tau_\mathrm{rise}$. For partial disruptions (where the lost mass is $\Delta M \lessapprox 0.8M_\star)$ the effect of SMBH's rotation on the $\dot{M}_\mathrm{peak}$ is reversed, because SMBH's rotation changes the amount of lost mass. For larger $\Delta M$ trend is similar. However, the effect of the SMBH's rotation is much smaller due to larger pericenter distances. We note that the dependencies of all the characteristics of fallback rate curves on the SMBH's spin follow a similar trend to $\beta$ dependency. 
	
	There have been only a few studies of the effect of SMBH's rotation on the mass fallback rate of the debris. The most extensive study, to our knowledge, is \citet{Gafton_2019}. They simulated disruptions of polytropic stars by a spinning SMBH and calculated characteristic properties of $\dot{M}$, such as $\dot{M}_\mathrm{peak}$, $t_\mathrm{peak}$, $t_\mathrm{Edd}$ and $n_\infty$. The general trend of the effect of SMBH's rotation is the same as found by us. $\dot{M}$ curves for different values of the SMBH's spin were also calculated by \citet{Kesden_2012}, who found that prograde spins reduce peak values of $\dot{M}$ and delay $t_\mathrm{peak}$. We recover the decrease in $\dot{M}_\mathrm{peak}$ in partial disruptions of TAMS stars. For high $\beta$ encounters we find the opposite trend.

 	The black hole's rotation can also delay the start of the self-crossing, the self-collision between the part of the stream falling towards the SMBH and the part receding from the SMBH, which drives the subsequent accretion disk formation \citep{liptai2019disc, Bonnerot_2020}. We stop our simulations before the most bound debris returns to the SMBH's vicinity and calculate $\dot{M}$ as the mass return rate at the moment of the second passage. In this way, we avoid the uncertainties due to the nozzle shock\footnote{During the pericenter passage, a strong vertical compression induced by an intersection of the inclined orbital planes of the returning gas results in a nozzle shock, which can change $\dot{M}$. } and the self-crossing. For non-rotating SMBHs, self-crossing occurs after the second passage, but before the third passage \citep{Hayasaki_2013, bonnerot}. For rotating black holes, the relativistic Lense-Thirring precession induces a change in the nodal angle of the angular momentum of the receding stream, which causes a misalignment between the colliding streams in the self-crossing region \citep{bonnerot_2020_book_arxiv}. In most extreme cases the streams can even miss each other and collide after several revolutions, mainly between adjacent orbital windings \citep{Batra_2022}. During each pericenter passage, $\dot{M}$ can change due to the nozzle shock. However, \citet{Jiang_2016} and \citet{bonnerot2021nozzle} find that the properties of the receding stream remain largely unaffected. This suggests that the mass fallback rate of the debris does not change significantly for adjacent windings and that our approach can be used to calculate $\dot{M}$ also at a later time provided that the streams do not collide by that time.

	\subsection{The effect of stellar and orbital parameters}
	
	$\dot{M}$ curves vary with $\beta$, stellar mass and age. Careful analysis of these dependencies and construction of the fitting functions allow a more accurate calculation of characteristic properties from the observed lightcurve. Mostly, we recover similar trends as previous studies such as \citet{Guillochon_2013, Ryu_2020b, lawsmith2020stellar}. The dependencies can be divided into two main regimes: partial and total disruptions.  These dependencies are directly related to the increase of the lost mass. Furthermore, the spread in energy is increasing with $\beta$, since the strength of the disruption increases as the pericenter distance decreases. 
	
	We confirm that the late-time slope in TTDEs is consistent with the theoretical prediction and follows $t^{-5/3}$. For PTDEs  \citet{Coughlin_2019} predicted a late-time decay $\propto t^{-9/4}$. We find a wide spread of power-law indices around the $n_\infty=-9/4$ value --- differences between the calculated $n_\infty$ and predicted value by \citet{Coughlin_2019} increase with the mass of the surviving stellar core.
	
	\subsubsection{Total disruptions}
	
	In total disruptions $\dot{M}_\mathrm{peak}$, $t_\mathrm{peak}$ and $\tau_\mathrm{rise}$ decrease, while $t_\mathrm{Edd}$ and $t_\mathrm{>0.5\dot{M}_\mathrm{peak}}$ increase with $\beta$ for a fixed stellar mass.  This is a consequence of a larger spread in energy of the debris as $\beta$ increases and stars are disrupted in a steeper tidal field. After the disruption the orbits of the bound debris span a wider range of Keplerian ellipses. The most bound debris returns sooner to the proximity of the SMBH, while the least bound debris returns later for higher values of $\beta$. This effect is visible in shorter $t_\mathrm{peak}$ and longer $t_\mathrm{Edd}$ and $t_\mathrm{>0.5\dot{M}_\mathrm{peak}}$. Because for TTDEs the integral of $\dot{M}$ over time needs to be constant, wider $\dot{M}$ results in lower  $\dot{M}_\mathrm{peak}$.

	At a fixed $\beta$, values of $\dot{M}_\mathrm{peak}$ and $t_\mathrm{Edd}$ increase, while the values of all the other characteristics decrease with $M_\star$ (or with $\rho_\mathrm{c}/\overline{\rho}$) in TTDEs. This can be understood from Equations (\ref{eq_dotMpeak}) and (\ref{eq_tedd}) --- for main sequence stars both quantities increase with $M_\star$. For $t_\mathrm{peak}$ Equation (\ref{eq_tpeak}) predicts an increase with $M_\star$ (for $M_\star \lessapprox 1.5\, \mathrm{M_\odot}$) and a decrease with $M_\star$ (for $M_\star \gtrapprox 1.5\, \mathrm{M_\odot}$), when a mass-radius relation is taken into account. However, we find that $t_\mathrm{peak}$ decreases in the entire $M_\star$ range in agreement with \citet{lawsmith2020stellar}. An increase in peak values of $\dot{M}$ and earlier peak times results in a shorter early rise-time scale. Furthermore, we find that $t_\mathrm{>0.5\dot{M}_\mathrm{peak}}$ decreases primarily with $\rho_\mathrm{c}/\overline{\rho}$ and not $M_\star$.
	
	\subsubsection{Partial disruptions}
	
	In partial disruptions the dependence on $\beta$, $M_\star$ and age is more nuanced --- there is a general trend with several exceptions. Since PTDEs are mostly disruptions of TAMS stars, the stellar outer layers are less bound than in disruptions of ZAMS stars. Therefore, the effect on $\dot{M}$ curves is determined by an interplay between the distance to the pericenter (determines the strength of the encounter) and the ratio $\rho_\mathrm{c}/\overline{\rho}$ (determines boundness of the outer layers).
	In order to more accurately determine how properties of $\dot{M}$ curves scale with stellar and orbital parameters it would be better to verify their dependence on a parameter, which would be a function of $\beta_\mathrm{crit}$ (when a TTDE occurs) and  $\rho_\mathrm{c}/\overline{\rho}$. \citet{lawsmith2020stellar} propose $\exp{(\beta/\beta_\mathrm{crit})^\alpha - 1}$, where $\alpha=(\rho_\mathrm{c}/\overline{\rho})^{-1/3}$. However, determining $\beta_\mathrm{crit}$ is beyond the scope of this work.

	$\dot{M}_\mathrm{peak}$, $t_\mathrm{Edd}$ and $t_\mathrm{>0.5\dot{M}_\mathrm{peak}}$ are increasing with $\beta$ at a fixed $M_\star$, while $t_\mathrm{peak}$ and $\tau_\mathrm{rise}$ are decreasing.  At a fixed parameter $\beta$, values of $\dot{M}_\mathrm{peak}$, $t_\mathrm{peak}$, $t_\mathrm{>0.5\dot{M}_\mathrm{peak}}$ and $\tau_\mathrm{rise}$ are increasing with $M_\star$, similar to total disruptions.

		\subsubsection{Physical tidal radius}
We determine parameter $\beta$ from the tidal radius $R_\mathrm{t} \propto \overline{\rho}^{-1/3}$, which does not depend on the stellar density profile. A more physical tidal radius $\mathcal{R}_\mathrm{t}\propto (\rho_\mathrm{c}/\overline{\rho})^{1/3}$ (for a fixed black hole mass), which determines the maximum pericenter distance for a total disruption, has been discussed in \citet{Ryu_2020a}. Therefore, it is possible to compare disruptions of different stars at the same physical parameter $\beta_\mathrm{p} = \mathcal{R}_\mathrm{t}/r_\mathrm{p}$ by scaling the pericenter distance with the compactness $ (\rho_\mathrm{c}/\overline{\rho})^{1/3}$. 

\begin{figure}
	\centering

					\includegraphics[width=\linewidth]{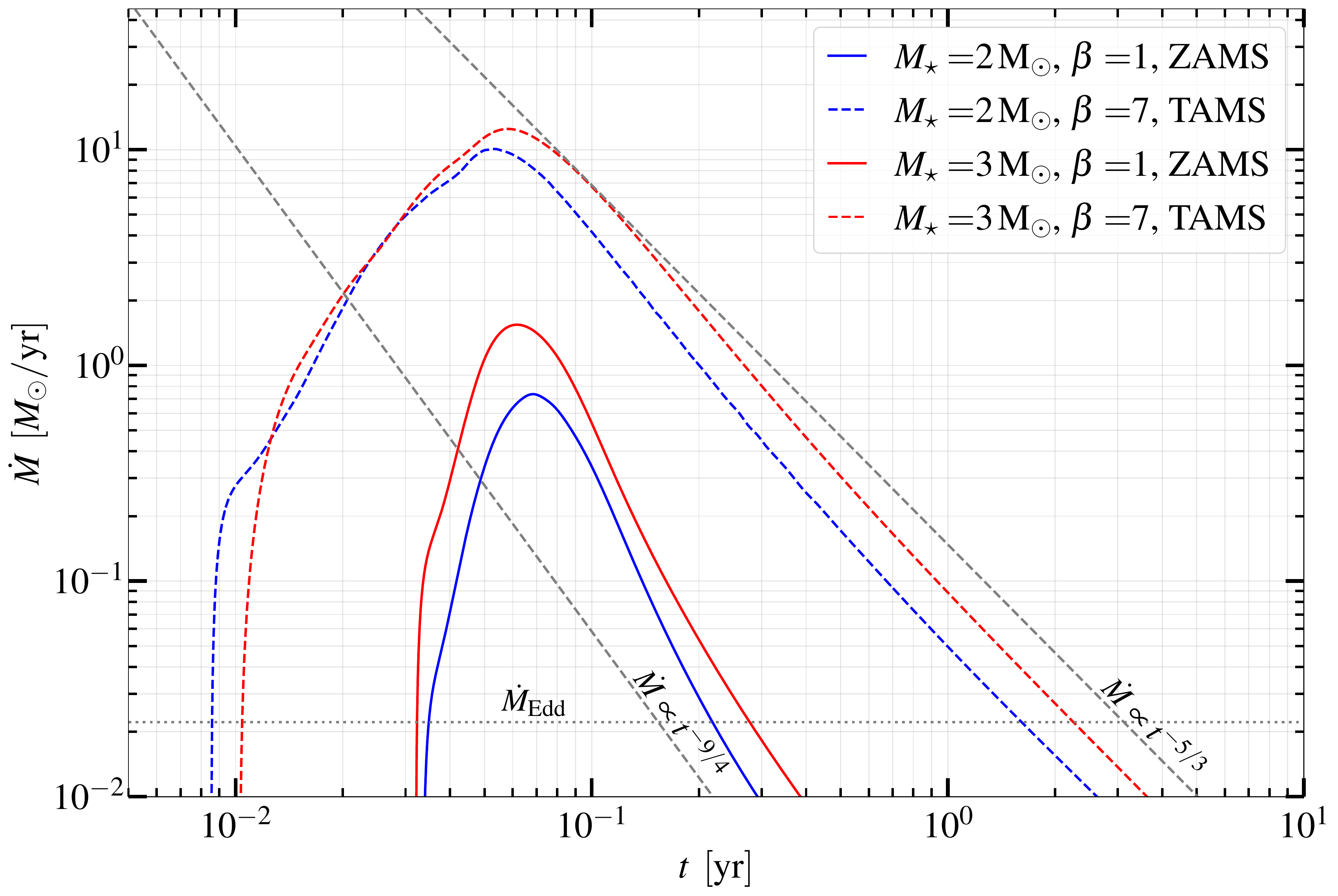}

	\caption{$\dot{M}$ for $2\, \mathrm{M_\odot}$ and $3\, \mathrm{M_\odot}$ ZAMS and TAMS stars at the same physical parameter $\beta_\mathrm{p}$ (indicated by the same colors). Horizontal dotted line indicates the Eddington accretion rate of a $10^6\, \mathrm{M_\odot}$ SMBH. Diagonal dotted lines represent power-law curves: $t^{-5/3}$ for total stellar disruptions, and $t^{-9/4}$ for partial stellar disruptions. }
	\label{fig:Rt_phys}
\end{figure}

From Table \ref{tab:ic} we estimate, that disruptions of $2\, \mathrm{M_\odot}$ and $3\, \mathrm{M_\odot}$ TAMS stars have, due to the higher compactness, shorter $\mathcal{R}_\mathrm{t}$ by a factor of $\approx 3$ in comparison to their ZAMS counterparts. Therefore, partial disruption of a $2\, \mathrm{M_\odot}$ ZAMS star on an orbit with $\beta=1$ has an equal $\beta_\mathrm{p}$ as partial disruption of a $2\, \mathrm{M_\odot}$ TAMS star on an orbit with $\beta=7$ (the same holds also for $3\, \mathrm{M_\odot}$ ZAMS star and $3\, \mathrm{M_\odot}$ TAMS star on an orbit with $\beta=1$ and $\beta=7$, respectively). In Figure \ref{fig:Rt_phys} we compare $\dot{M}$ for these disruptions and find, that encounters of ZAMS stars result in a lower mass loss, while disruptions of TAMS stars are very close to TTDEs, contrary to results presented in Section \ref{sec3}. We see, that even though disruptions are scaled to the same $\beta_\mathrm{p}$, the outcome is very different.

We also compare $\dot{M}$ from a total disruption of a $2\, \mathrm{M_\odot}$ ZAMS star on an orbit with $\beta=3$ and a total disruption of a $2\, \mathrm{M_\odot}$ TAMS star on an orbit with $\beta=18.1$, where both encounters have an equal $\beta_\mathrm{p}$. We find, that the outcome of the interaction between the SMBH and the star is substantially different already during the first passage. In the first case, there is no debris moving on plunging orbits, while in the second case, $\approx 10\%$ of the total stellar mass plunges in the SMBH. From these examples we conclude, that $\beta$ and $\beta_\mathrm{p}$ cannot be used as clear indicators of an outcome of a TDE.

	\subsection{Potential caveats} \label{sec:caveats}
	
	Accurate self-gravity description is important in the treatment of various astrophysical phenomena. It is especially important in TDEs, e.g. in the calculations of $\dot{M}$ because it affects the distribution of the debris mass over total energy after the disruption. 
	
	The implementation of the self-gravity, the gravity between individual gas particles, is not trivial in relativistic simulations. In order to calculate the total force due to the self-gravity, experienced by a certain particle, it is necessary to determine the contribution from all other particles. Therefore, an accurate GR simulation would require Einstein’s equations to be solved for each particle, which is not feasible with the current computational technology. In order to circumvent this problem, self-gravity is described in a Newtonian way or in a GR approximation \citep{Gafton_2019, Ryu_2020b}. 
	
	In our simulations we use a Newtonian self-gravity, while \citet{Gafton_2019} and \citet{Ryu_2020b} used different relativistic approximations. The research by \citet{Gafton_2019} is especially interesting, since their study also addressed relativistic effects on $\dot{M}$ over a wide range of $\beta$. However, their results differ quantitatively as well as qualitatively from ours. This could be due to the only notable difference between their approach and ours, which we could identify --- a different implementation of self-gravity\footnote{We note that they used a different SPH code and that they studied disruptions of a polytropic star. However, this does not explain the qualitative differences between our results.}. Currently there is no method available, which would test the validity of the self-gravity implementation. The development of such a method is beyond the scope of this paper.
	
	Another potential caveat is related to the conversion of the density profile obtained with \textsc{MESA} to a 3D density distribution of particles with the program \textsc{MESA2HYDRO}. This conversion results in a higher density in the stellar center due to the presence of a sink particle. Comparing values of $\rho_\mathrm{c}/\overline{\rho}$ from Table \ref{tab:ic} to the values from Table 1 in \citet{lawsmith2020stellar} we see some discrepancies. \citet{Joyce_2019} speculated that these discrepancies should not have any major consequences on the process of disruption. We confirm this by comparing several $\dot{M}$ from NR disruptions to $\dot{M}$ curves constructed with the \textsc{STARS} library \citep{lawsmith2020stellar}.

	\section{Summary} \label{sec5}
	
	We calculate mass fallback rate of the debris $\dot{M}$ in a Newtonian and a general relativistic gravitational potential of a $10^6\, \mathrm{M_\odot}$ SMBH for different stellar masses $M_\star$, ages, $\beta$ parameter and SMBH's spins. We calculate peak values $\dot{M}_\mathrm{peak}$, time to the peak $t_\mathrm{peak}$, duration of the super-Eddington phase $t_{>\dot{M}_\mathrm{Edd}}$, duration $t_{>0.5\dot{M}_\mathrm{peak}}$, characteristic rise-time $\tau_\mathrm{rise}$ and late-time slope $n_\infty$. Summary of our main results is:

	\begin{itemize}
		\item  We recover the trends of $\dot{M}_\mathrm{peak}$, $t_\mathrm{peak}$, $t_\mathrm{Edd}$, $\tau_\mathrm{rise}$ and $n_\infty$ with $\beta$, stellar mass and age, which were obtained in previous studies. We also find that the trends can change for stars with the mass  $\approx 3\,\mathrm{M_\odot}$ stars due to the convective mixing, which decreases the central density.
		\item At a fixed $\beta$ and stellar age, $t_\mathrm{Edd}$ increases with $M_\star$, while $t_\mathrm{>0.5\dot{M}_\mathrm{peak}}$ increases with the compactness of stars $\rho_\mathrm{c}/\overline{\rho}$.  
  \item Comparison of $\dot{M}$ from disruptions of stars (with the same $M_\star$ and different age) at an equal parameter $\beta\propto \overline{\rho}^{-1/3}$ or physical parameter $\beta_\mathrm{p}\propto (\rho_\mathrm{c}/\overline{\rho})^{1/3}$  is not a clear indicator of an outcome of a TDE.
		\item Disruptions at the same pericenter distance in a relativistic tidal field are stronger than in a Newtonian. This results in relativistic $\dot{M}$ curves with higher $\dot{M}_\mathrm{peak}$, and lower values of $t_\mathrm{peak}$, $\tau_\mathrm{rise}$, $t_\mathrm{Edd}$ and $t_\mathrm{>0.5\dot{M}_\mathrm{peak}}$.
		\item Differences between a relativistic treatment of the SMBH's gravity and a non-relativistic are apparent even for $\beta=1$ encounters. This emphasizes the importance to treat SMBH's gravity with a general relativistic description.
		\item For stars on prograde orbits, rotation of the SMBH results in a similar effect on $\dot{M}$ as decreasing the pericenter distance, while the opposite happens for stars on retrograde orbits.
	\end{itemize}

%% Also note that the akcnowlodgment environment does not support long amounts of text. If you have a lot of people and institutions to acknowledge, do not use this command. Instead, create a new \section{Acknowledgments}.
\begin{acknowledgments}
	This research was supported by the Slovenian Research Agency grants P1-0031, I0-0033, J1-8136, J1-2460 and the Young Researchers program. We acknowledge
	the use of \textsc{MESA}, \textsc{MESA2HYDRO} and \textsc{Phantom} for detailed hydrodynamical simulations and \textsc{SPLASH} for the visualization of the output \citep{Paxton_2010, Joyce_2019, price}. We gratefully acknowledge the HPC RIVR consortium (\href{www.hpc-rivr.si}{www.hpc-rivr.si}) and EuroHPC JU (\href{eurohpc-ju.europa.eu}{eurohpc-ju.europa.eu}) for funding this research by providing computing resources of the HPC system Vega at the Institute of Information Science (\href{www.izum.si}{www.izum.si}).  We thank M. Joyce and E. Tejeda for the useful discussions. We are grateful to the anonymous referee for the comments and suggestions, which helped us improve the paper.
	
\end{acknowledgments}





\appendix

		\section{Parameters of simulations} \label{app:sim}
	Relevant parameters used in \textsc{MESA}, \textsc{MESA2HYDRO} and \textsc{Phantom} are shown in Tables \ref{tab:ic_mesa}, \ref{tab:ic_mesa2} and \ref{tab:ic_ph}, respectively. \textsc{MESA} \emph{inlist} files are available on Zenodo: \dataset[10.5281/zenodo.7428262]{https://doi.org/10.5281/zenodo.7428262}.
	
	\begin{table}
		\caption{Relevant \textsc{MESA} parameters. For stars with masses $M_\star > 1.2\,\mathrm{M_\odot}$ we also take into account overshooting.}             % title of Table
		\label{tab:ic_mesa}      % is used to refer this table in the text
		\centering                          % used for centering table
		\begin{tabular}{l | l }        % centered columns
			\hline\hline                 % inserts double horizontal lines
			Parameter  & Value   \\    % table heading 
			\hline\hline                      % inserts single horizontal line
			$\mathrm{create\_pre\_main\_sequence\_model}$ & .true.  \\
			$\mathrm{new\_net\_name}$ & '$\mathrm{mesa\_49.net}$'  \\
			$\mathrm{new\_rate\_preference}$ &  $\mathrm{2 }$ !jina \\
			$\mathrm{kappa\_file\_prefix}$ & '$\mathrm{09}$'  \\
			$\mathrm{initial\_zfracs }$ &  $\mathrm{ 6}$ \\
			$\mathrm{kappa\_lowT\_prefix}$ &  '$\mathrm{lowT\_fa05\_a09p}$' \\
			$\mathrm{initial\_z}$ &  $\mathrm{0.0142}$ \\
			$\mathrm{initial\_y}$ & $\mathrm{0.2703}$  \\
			$\mathrm{Lnuc\_div\_L\_zams\_limit}$ &  $\mathrm{0.999}$ \\
			$\mathrm{mixing\_length\_alpha}$ & $\mathrm{2}$  \\
			$\mathrm{delta\_lg\_XH\_cntr\_hard\_limit}$ & $\mathrm{0.005}$  \\
			$\mathrm{do\_element\_diffusion}$ &  $\mathrm{.true.}$ \\
			$\mathrm{xa\_central\_lower\_limit species(1)}$ &  '$\mathrm{h1}$' \\
			$\mathrm{xa\_central\_lower\_limit(1)}$ &  $\mathrm{0.001}$ \\
			$\mathrm{maxovershoot\_scheme(1)}$ &  'exponential'\\
			$\mathrm{overshoot\_zone\_type(1)}$ &  'any' \\
			$\mathrm{overshoot\_zone\_loc(1)}$ &  'any' \\
			$\mathrm{overshoot\_bdy\_loc(1)}$ &  'any' \\
			$\mathrm{overshoot\_f(1)}$ &  0.014 \\
			$\mathrm{overshoot\_f0(1)}$ &  0.004 \\
			\hline                                   %inserts single line
		\end{tabular}
	\end{table}

	\begin{table}
		\caption{Relevant \textsc{MESA2HYDRO} parameters.}             % title of Table
		\label{tab:ic_mesa2}      % is used to refer this table in the text
		\centering                          % used for centering table
		\begin{tabular}{l | l }        % centered columns
			\hline\hline                 % inserts double horizontal lines
			Parameter  & Value   \\    % table heading 
			\hline\hline                      % inserts single horizontal line
			$\mathrm{IC\_format\_type}$ & $\mathrm{phantom\_binary}$  \\
			$\mathrm{r\_depth}$ &  $1\mathrm{e}-20$ \\
			$\mathrm{N}$ & 8  \\
			$\mathrm{mp}$ & $1\mathrm{e-6}$  \\
			$\mathrm{stepsize}$ & $1\mathrm{e}4$  \\
			$\mathrm{which\_dtype=}$ & d  \\
			$\mathrm{TOL}$ & 0.01  \\
			\hline                                   %inserts single line
		\end{tabular}
	\end{table}

	\begin{table}
		\caption{Relevant \textsc{Phantom} parameters. The softening length $\mathrm{h\_soft\_sinksink}$ of the sink particle is calibrated to be half of the the inner-most radius of the gas shells. We also use parameters $\mathrm{GRAVITY=yes}$, $\mathrm{ISOTHERMAL=no}$ and, for GR simulations, $\mathrm{GR=yes}$ and $\mathrm{METRIC=kerr}$.}             % title of Table
		\label{tab:ic_ph}      % is used to refer this table in the text
		\centering                          % used for centering table
		\begin{tabular}{l | l }        % centered columns
			\hline\hline                 % inserts double horizontal lines
			Parameter  & Value   \\    % table heading 
			\hline\hline                      % inserts single horizontal line
			$\mathrm{alpha}$ & 1  \\
			$\mathrm{alphau}$ &  0.1 \\
			$\mathrm{beta}$ & 1 \\
			$\mathrm{ieos}$ & 2 \\
			$\mathrm{h\_soft\_sinksink}$ & 0.01188 (for $1\,\mathrm{M_\odot}$, ZAMS star) \\
			$\mathrm{accradius1}$ & 5  \\
			$\mathrm{mass1}$ &  1 \\
			$\mathrm{a}$ & 0. or 0.99  \\
%				\hline                exist                   %inserts single line
			\end{tabular}
		\end{table}

		\section{Resolution test} \label{app:1}

\begin{figure} [htb!]
			\centering
			
			\includegraphics[width=0.69\textwidth]{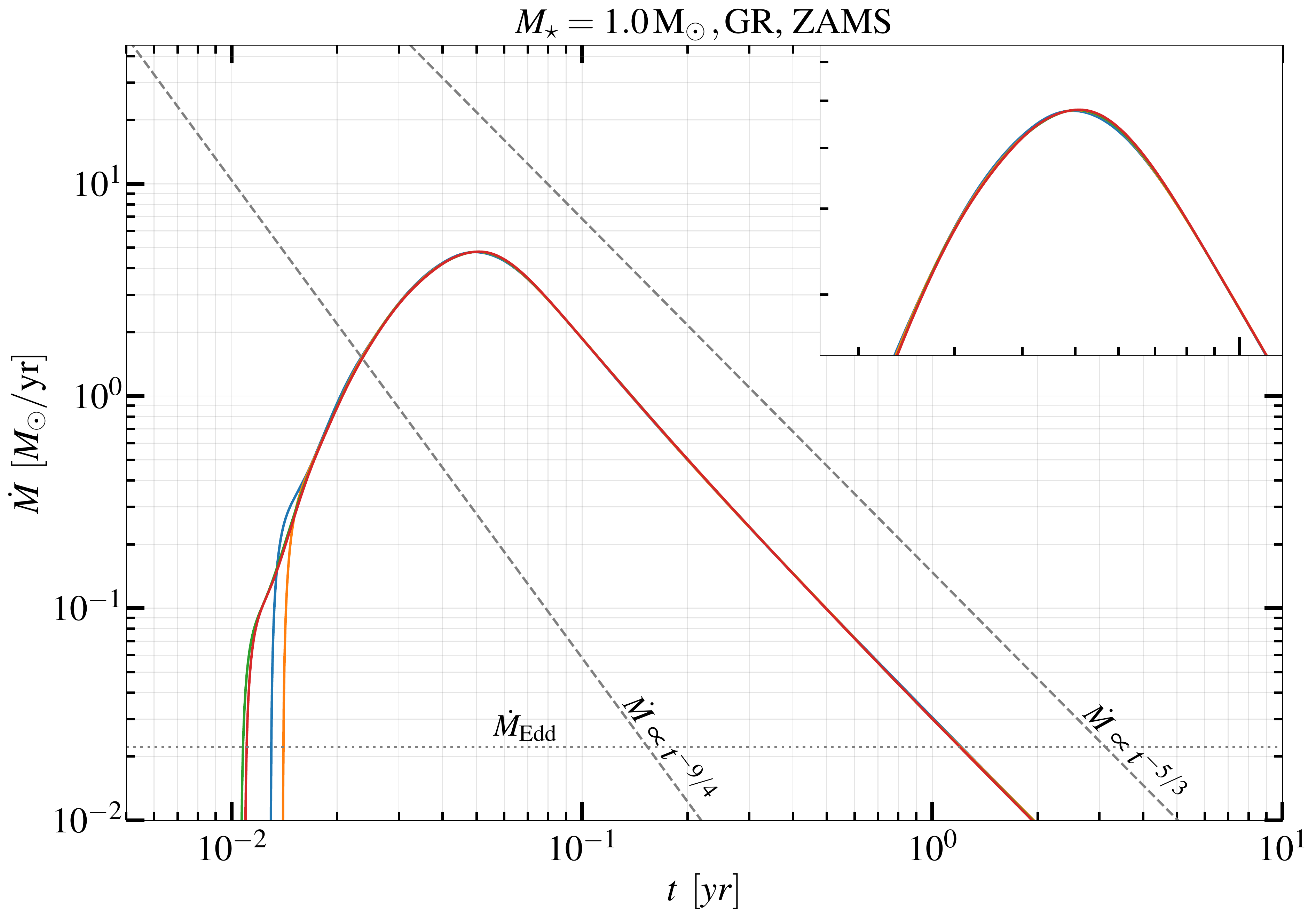}
			
			\begin{minipage}[b]{\linewidth}
				\centering
				\includegraphics[width=0.52\textwidth]{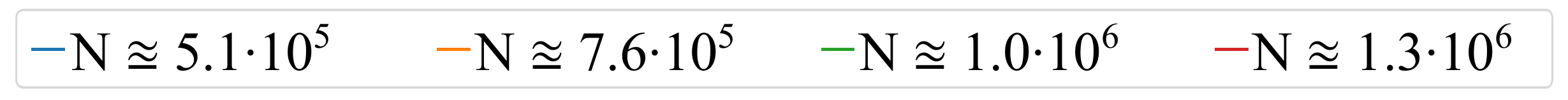}
			\end{minipage}
			\vspace*{-0.1cm}
			
			\caption{$\dot{M}$ for $1\, \mathrm{M_\odot}$ ZAMS stars disrupted by a non-rotating SMBH for $\beta=3$ and number of particles $N\approxeq 5.1\cdot10^5$, $7.6\cdot10^5$, $1.0\cdot10^6$, $1.3\cdot10^6$. Horizontal dotted line indicates the Eddington accretion rate of a $10^6\, \mathrm{M_\odot}$ SMBH. Diagonal dotted lines represent power-law curves: $t^{-5/3}$ for total stellar disruptions, and $t^{-9/4}$ for partial stellar disruptions.}
			\label{s:a1}
		\end{figure}

		We have performed a resolution test to determine if the results converge with the increasing number of particles $N$. We have performed simulations with $N\approxeq 5.1\cdot10^5$, $7.6\cdot10^5$, $1.0\cdot10^6$, $1.3\cdot10^6$. The results are shown in Figure \ref{s:a1}. In the range of $\dot{M} \gtrapprox 0.05\dot{M}_\mathrm{peak}$ we find no noticeable differences between the results for different numbers of particles. There are minor discrepancies for low values of $t$ for $\dot{M}\lessapprox  0.05\dot{M}_\mathrm{peak}$, at the order of $< 0.04 t_\mathrm{peak}$. Since $t_\mathrm{peak}$ is the shortest evaluated time scale in our study, we conclude that our results are sufficiently accurate.

\newpage
\bibliography{cite}{}

\begin{thebibliography}{}
\expandafter\ifx\csname natexlab\endcsname\relax\def\natexlab#1{#1}\fi
\providecommand{\url}[1]{\href{#1}{#1}}
\providecommand{\dodoi}[1]{doi:~\href{http://doi.org/#1}{\nolinkurl{#1}}}
\providecommand{\doeprint}[1]{\href{http://ascl.net/#1}{\nolinkurl{http://ascl.net/#1}}}
\providecommand{\doarXiv}[1]{\href{https://arxiv.org/abs/#1}{\nolinkurl{https://arxiv.org/abs/#1}}}

\bibitem[{Alexander(2017)}]{Alexander_2017}
Alexander, T. 2017, Annual Review of Astronomy and Astrophysics, 55, 17–57,
  \dodoi{10.1146/annurev-astro-091916-055306}

\bibitem[{Ayal {et~al.}(2000)Ayal, Livio, \& Piran}]{Ayal_2000}
Ayal, S., Livio, M., \& Piran, T. 2000, The Astrophysical Journal, 545, 772,
  \dodoi{10.1086/317835}

\bibitem[{Batra {et~al.}(2021)Batra, Lu, Bonnerot, \& Phinney}]{Batra_2022}
Batra, G., Lu, W., Bonnerot, C., \& Phinney, E.~S. 2021, General Relativistic
  Stream Crossing in Tidal Disruption Events,  arXiv,
  \dodoi{10.48550/ARXIV.2112.03918}

\bibitem[{Bonnerot \& Lu(2020)}]{Bonnerot_2020}
Bonnerot, C., \& Lu, W. 2020, Monthly Notices of the Royal Astronomical
  Society, \dodoi{10.1093/mnras/staa1246}

\bibitem[{Bonnerot \& Lu(2021)}]{bonnerot2021nozzle}
---. 2021, The nozzle shock in tidal disruption events.
\newblock \doarXiv{2106.01376}

\bibitem[{Bonnerot {et~al.}(2015)Bonnerot, Rossi, Lodato, \& Price}]{bonnerot}
Bonnerot, C., Rossi, E.~M., Lodato, G., \& Price, D.~J. 2015, Monthly Notices
  of the Royal Astronomical Society, 455, 2253, \dodoi{10.1093/mnras/stv2411}

\bibitem[{Bonnerot \& Stone(2020)}]{bonnerot_2020_book_arxiv}
Bonnerot, C., \& Stone, N. 2020, Formation of an Accretion Flow,  arXiv,
  \dodoi{10.48550/ARXIV.2008.11731}

\bibitem[{Bricman \& Gomboc(2020)}]{Bricman_2020}
Bricman, K., \& Gomboc, A. 2020, The Astrophysical Journal, 890, 73,
  \dodoi{10.3847/1538-4357/ab6989}

\bibitem[{Carter \& Luminet(1982)}]{Carter_1982}
Carter, B., \& Luminet, J.~P. 1982, Nature, 296, 211, \dodoi{10.1038/296211a0}

\bibitem[{{Carter} \& {Luminet}(1985)}]{Carter_1985}
{Carter}, B., \& {Luminet}, J.~P. 1985, \mnras, 212, 23,
  \dodoi{10.1093/mnras/212.1.23}

\bibitem[{{Cheng} \& {Bogdanovi{\'c}}(2014)}]{Cheng_2014}
{Cheng}, R.~M., \& {Bogdanovi{\'c}}, T. 2014, \prd, 90, 064020,
  \dodoi{10.1103/PhysRevD.90.064020}

\bibitem[{Clerici \& Gomboc(2020)}]{Clerici_Gomboc_2020}
Clerici, A., \& Gomboc, A. 2020, Astronomy \& Astrophysics, 642,
  \dodoi{10.1051/0004-6361/202037641}

\bibitem[{Coughlin \& Nixon(2019)}]{Coughlin_2019}
Coughlin, E.~R., \& Nixon, C.~J. 2019, The Astrophysical Journal, 883, L17,
  \dodoi{10.3847/2041-8213/ab412d}

\bibitem[{{Demircan} \& {Kahraman}(1991)}]{m_r_paper_1991}
{Demircan}, O., \& {Kahraman}, G. 1991, \apss, 181, 313,
  \dodoi{10.1007/BF00639097}

\bibitem[{Evans \& Kochanek(1989)}]{evans}
Evans, C.~R., \& Kochanek, C.~S. 1989, The Astrophysical Journal, 346, L13,
  \dodoi{10.1086/185567}

\bibitem[{Gafton \& Rosswog(2019)}]{Gafton_2019}
Gafton, E., \& Rosswog, S. 2019, Monthly Notices of the Royal Astronomical
  Society, 487, 4790–4808, \dodoi{10.1093/mnras/stz1530}

\bibitem[{Golightly {et~al.}(2019)Golightly, Nixon, \&
  Coughlin}]{Golightly_2019}
Golightly, E. C.~A., Nixon, C.~J., \& Coughlin, E.~R. 2019, The Astrophysical
  Journal, 882, L26, \dodoi{10.3847/2041-8213/ab380d}

\bibitem[{Guillochon \& Ramirez-Ruiz(2013)}]{Guillochon_2013}
Guillochon, J., \& Ramirez-Ruiz, E. 2013, The Astrophysical Journal, 767, 25,
  \dodoi{10.1088/0004-637x/767/1/25}

\bibitem[{Hayasaki {et~al.}(2013)Hayasaki, Stone, \& Loeb}]{Hayasaki_2013}
Hayasaki, K., Stone, N., \& Loeb, A. 2013, Monthly Notices of the Royal
  Astronomical Society, 434, 909–924, \dodoi{10.1093/mnras/stt871}

\bibitem[{Jiang {et~al.}(2016)Jiang, Guillochon, \& Loeb}]{Jiang_2016}
Jiang, Y.-F., Guillochon, J., \& Loeb, A. 2016, The Astrophysical Journal, 830,
  125, \dodoi{10.3847/0004-637x/830/2/125}

\bibitem[{Joyce {et~al.}(2019)Joyce, Lairmore, Price, Mohamed, \&
  Reichardt}]{Joyce_2019}
Joyce, M., Lairmore, L., Price, D.~J., Mohamed, S., \& Reichardt, T. 2019, The
  Astrophysical Journal, 882, 63, \dodoi{10.3847/1538-4357/ab3405}

\bibitem[{Kesden(2012)}]{Kesden_2012}
Kesden, M. 2012, Phys. Rev. D, 86, 064026, \dodoi{10.1103/PhysRevD.86.064026}

\bibitem[{Komossa(2015)}]{komossa}
Komossa, S. 2015, Journal of High Energy Astrophysics, 7, 148,
  \dodoi{10.1016/j.jheap.2015.04.006}

\bibitem[{Lamers \& M.~Levesque(2017)}]{book_sa2}
Lamers, H.~J., \& M.~Levesque, E. 2017, Understanding Stellar Evolution,
  2514-3433 (IOP Publishing), \dodoi{10.1088/978-0-7503-1278-3}

\bibitem[{Law-Smith {et~al.}(2019)Law-Smith, Guillochon, \&
  Ramirez-Ruiz}]{Law_Smith_2019}
Law-Smith, J., Guillochon, J., \& Ramirez-Ruiz, E. 2019, The Astrophysical
  Journal, 882, L25, \dodoi{10.3847/2041-8213/ab379a}

\bibitem[{Law-Smith {et~al.}(2020)Law-Smith, Coulter, Guillochon, Mockler, \&
  Ramirez-Ruiz}]{lawsmith2020stellar}
Law-Smith, J. A.~P., Coulter, D.~A., Guillochon, J., Mockler, B., \&
  Ramirez-Ruiz, E. 2020, Stellar TDEs with Abundances and Realistic Structures
  (STARS): Library of Fallback Rates.
\newblock \doarXiv{2007.10996}

\bibitem[{Liptai {et~al.}(2019)Liptai, Price, Mandel, \&
  Lodato}]{liptai2019disc}
Liptai, D., Price, D.~J., Mandel, I., \& Lodato, G. 2019, Disc formation from
  tidal disruption of stars on eccentric orbits by Kerr black holes using
  GRSPH.
\newblock \doarXiv{1910.10154}

\bibitem[{Lodato {et~al.}(2009)Lodato, King, \& Pringle}]{Lodato_2009}
Lodato, G., King, A.~R., \& Pringle, J.~E. 2009, Monthly Notices of the Royal
  Astronomical Society, 392, 332, \dodoi{10.1111/j.1365-2966.2008.14049.x}

\bibitem[{Paxton {et~al.}(2010)Paxton, Bildsten, Dotter, Herwig, Lesaffre, \&
  Timmes}]{Paxton_2010}
Paxton, B., Bildsten, L., Dotter, A., {et~al.} 2010, The Astrophysical Journal
  Supplement Series, 192, 3, \dodoi{10.1088/0067-0049/192/1/3}

\bibitem[{Price \& et. al.(2018)}]{price}
Price, D.~J., \& et. al. 2018, Publications of the Astronomical Society of
  Australia, 35, \dodoi{10.1017/pasa.2018.25}

\bibitem[{Rees(1988)}]{rees}
Rees, M.~J. 1988, Nature, 333, 523, \dodoi{10.1038/333523a0}

\bibitem[{Ryu {et~al.}(2020{\natexlab{a}})Ryu, Krolik, \& Piran}]{Ryu_2020a}
Ryu, T., Krolik, J., \& Piran, T. 2020{\natexlab{a}}, The Astrophysical
  Journal, 904, 73, \dodoi{10.3847/1538-4357/abbf4d}

\bibitem[{Ryu {et~al.}(2020{\natexlab{b}})Ryu, Krolik, Piran, \&
  Noble}]{Ryu_2020b}
Ryu, T., Krolik, J., Piran, T., \& Noble, S.~C. 2020{\natexlab{b}}, The
  Astrophysical Journal, 904, 98, \dodoi{10.3847/1538-4357/abb3cf}

\bibitem[{Ryu {et~al.}(2020{\natexlab{c}})Ryu, Krolik, Piran, \&
  Noble}]{Ryu_2020c}
---. 2020{\natexlab{c}}, The Astrophysical Journal, 904, 100,
  \dodoi{10.3847/1538-4357/abb3ce}

\bibitem[{Ryu {et~al.}(2020{\natexlab{d}})Ryu, Krolik, Piran, \&
  Noble}]{Ryu_2020d}
---. 2020{\natexlab{d}}, The Astrophysical Journal, 904, 101,
  \dodoi{10.3847/1538-4357/abb3cc}

\bibitem[{{Servin} \& {Kesden}(2017)}]{Servin_2017}
{Servin}, J., \& {Kesden}, M. 2017, \prd, 95, 083001,
  \dodoi{10.1103/PhysRevD.95.083001}

\bibitem[{Stone(2015)}]{Stone2015}
Stone, N.~C. 2015, The Tidal Disruption of Stars by Supermassive Black Holes
  (Springer International Publishing), \dodoi{10.1007/978-3-319-12676-0}

\bibitem[{Stone {et~al.}(2019)Stone, Kesden, Cheng, \& van Velzen}]{Stone_2019}
Stone, N.~C., Kesden, M., Cheng, R.~M., \& van Velzen, S. 2019, General
  Relativity and Gravitation, 51, \dodoi{10.1007/s10714-019-2510-9}

\bibitem[{Tejeda {et~al.}(2017)Tejeda, Gafton, Rosswog, \&
  Miller}]{Tejeda_2017}
Tejeda, E., Gafton, E., Rosswog, S., \& Miller, J.~C. 2017, Monthly Notices of
  the Royal Astronomical Society, 469, 4483–4503,
  \dodoi{10.1093/mnras/stx1089}

\bibitem[{Tejeda \& Rosswog(2013)}]{tejeda}
Tejeda, E., \& Rosswog, S. 2013, Monthly Notices of the Royal Astronomical
  Society, 433, 1930, \dodoi{10.1093/mnras/stt853}

\bibitem[{van Velzen \& et~al.(2019)}]{van_Velzen_2019}
van Velzen, S., \& et~al. 2019, The Astrophysical Journal, 872, 198,
  \dodoi{10.3847/1538-4357/aafe0c}

\bibitem[{van Velzen {et~al.}(2011)van Velzen, Farrar, Gezari, Morrell,
  Zaritsky, {\"O}stman, Smith, Gelfand, \& Drake}]{van_Velzen_2011}
van Velzen, S., Farrar, G.~R., Gezari, S., {et~al.} 2011, The Astrophysical
  Journal, 741, 73, \dodoi{10.1088/0004-637x/741/2/73}

\bibitem[{van Velzen {et~al.}(2021)van Velzen, Gezari, Hammerstein, Roth,
  Frederick, Ward, Hung, Cenko, Stein, Perley, Taggart, Foley, Sollerman,
  Blagorodnova, Andreoni, Bellm, Brinnel, De, Dekany, Feeney, Fremling, Giomi,
  Golkhou, Graham, Ho, Kasliwal, Kilpatrick, Kulkarni, Kupfer, Laher, Mahabal,
  Masci, Miller, Nordin, Riddle, Rusholme, van Santen, Sharma, Shupe, \&
  Soumagnac}]{van_Velzen_2021}
van Velzen, S., Gezari, S., Hammerstein, E., {et~al.} 2021, The Astrophysical
  Journal, 908, 4, \dodoi{10.3847/1538-4357/abc258}

\end{thebibliography}
\bibliographystyle{aasjournal}



\end{document}